\documentclass{article}

\usepackage{amstext,amsmath,amssymb}
\usepackage[dvips]{color}
\usepackage{graphicx}
\usepackage{epsfig}
\usepackage{cite}

\begin{document}

\numberwithin{equation}{section}

\setcounter{secnumdepth}{2}

\bibliographystyle{unsrt}

\title{Quantum mechanical calculation of the effects of stiff and rigid constraints in the conformational equilibrium of the Alanine dipeptide}

\author{Pablo Echenique$^{1,2}$\footnote{Corresponding author. E-mail
address: {\tt pnique@unizar.es}}\ , Iv\'{a}n Calvo$^{1,2}$ and
J. L. Alonso$^{1,2}$
\vspace{0.4cm}\\ $^{1}$ {\small Departamento de F\'{\i}sica
Te\'orica, Facultad de Ciencias, Universidad de Zaragoza,}\\
{\small Pedro Cerbuna 12, 50009, Zaragoza, Spain.}\\
$^{2}$ {\small Instituto de Biocomputaci\'on y
F\'{\i}sica de los Sistemas Complejos (BIFI),}\\ {\small Edificio
Cervantes, Corona de Arag\'on 42, 50009, Zaragoza, Spain.}}

\date{\today}

\maketitle

\begin{abstract}

If constraints are imposed on a macromolecule, two inequivalent
classical models may be used: the stiff and the rigid one. This work
studies the effects of such constraints on the Conformational
Equilibrium Distribution (CED) of the model dipeptide HCO-{\small
L}-Ala-NH$_2$ \emph{without any simplifying assumption}. We use ab
initio Quantum Mechanics calculations including electron correlation
at the MP2 level to describe the system, and we measure the
conformational dependence of all the correcting terms to the naive CED
based in the Potential Energy Surface (PES) that appear when the
constraints are considered. These terms are related to mass-metric
tensors determinants and also occur in the Fixman's compensating
potential. We show that some of the corrections are non-negligible if
one is interested in the whole Ramachandran space. On the other hand,
if only the energetically lower region, containing the principal
secondary structure elements, is assumed to be relevant, then, all
correcting terms may be neglected up to peptides of considerable
length. This is the first time, as far as we know, that the analysis
of the conformational dependence of these correcting terms is
performed in a relevant biomolecule with a realistic potential energy
function.
\vspace{0.2cm}
\vspace{0.2cm}\\

\end{abstract}

\section{Introduction}
\label{sec:introduction}

In computer simulations of large complex systems, such as
macromolecules and, specially, proteins
\cite{PE:Alo2004BOOK,PE:Dob2003NAT,PE:Dil1999PSC,PE:He1998JCPb,PE:Aba1994JCC,PE:VGu1982MM},
one of the main bottlenecks to design efficient algorithms is the
necessity to sample an astronomically large conformational space
\cite{PE:Dil1999PSC,PE:Lev1969PROC}. In addition, being the typical
timescales of the different movements in a wide range, demandingly
small timesteps must be used in Molecular Dynamics simulations in
order to properly account for the fastest modes, which lie in the
femtosecond range. However, most of the biological interesting
behaviour (allosteric transitions, protein folding, enzymatic
catalysis) is related to the slowest conformational changes, which
occur in the timescale of milliseconds or even seconds
\cite{PE:Chu2000JCC,PE:Rei2000PD,PE:Rei1999JCOP,PE:He1998JCPb,PE:Sch1997ARBBS}.
Fortunately, the fastest modes are also the most energetic ones and
are rarely activated at room temperature. Therefore, in order to
alleviate the computational problems and also simplify the images used
to think about these elusive systems, one may naturally consider the
reduction of the number of degrees of freedom describing
macromolecules via the imposition of constraints \cite{PE:VKa1984AJP}.

How to study the conformational equilibrium of these constrained
systems has been an object of much debate
\cite{PE:Ral1979JFM,PE:Hel1979JCP,PE:Go1976MM,PE:Fix1974PNAS,PE:Go1969JCP}.
Two different classical models exist in the literature which are
conceptually
\cite{PE:Mor2004ACP,PE:Den2000MP,PE:Ral1979JFM,PE:Hel1979JCP,PE:Go1976MM,PE:Fix1974PNAS}
and practically
\cite{PE:Pat2004JCP,PE:Per1985MM,PE:VGu1982MM,PE:Ral1979JFM,PE:Pea1979JCP,PE:Cha1979JCP,PE:Got1976JCP}
inequivalent. In the \emph{classical rigid} model, the constraints are
assumed to be \emph{exact} and all the velocities that are orthogonal
to the hypersurface defined by them vanish. In the \emph{classical
stiff}\footnote{\label{foot:stiff_vs_flexible}Some authors use the
word \emph{flexible} to refer to this model
\cite{PE:Zho2000JCP,PE:Per1985MM,PE:Pea1979JCP,PE:Go1976MM}. We,
however, prefer to term it \emph{stiff} \cite{PE:Mor2004ACP} and keep
the name \emph{flexible} to refer to the case in which no constraints
are imposed.} model, on the other hand, the constraints are assumed to
be \emph{approximate} and they are implemented by a steep potential
that drives the system to the constrained hypersurface. In this case,
the orthogonal velocities are activated and may act as ``heat
containers''.

In this work, we do not address the question of which model is a
better approximation of physical reality. Although, in the literature,
it is commonly assumed (often implicitly) that the classical stiff
model should be taken as a reference
\cite{PE:Pat2004JCP,PE:Rei2000PD,PE:Den2000MP,PE:Ber1983BOOK,PE:VGu1982MM,PE:Pea1979JCP,PE:Fix1974PNAS},
we believe that this opinion is much influenced by the use of popular
classical force fields
\cite{PE:Mac1998BOOK,PE:Bro1983JCC,PE:VGu1982MM,PE:Cor1995JACS,PE:Pea1995CPC,PE:Jor1988JACS,PE:Jor1996JACS,PE:Hal1996JCCa,PE:Hal1996JCCb,PE:Hal1996JCCc,PE:Hal1996JCCd,PE:Hal1996JCCe}
(which are stiff by construction) and by the goal of reproducing their
results at a lower computational cost, i.e., using rigid Molecular
Dynamics simulations
\cite{PE:Pas2002JCP,PE:Chu2000JCC,PE:Rei2000PD,PE:Zho2000JCP,PE:Den2000MP,PE:Den1998JCP,PE:He1998JCPb,PE:Aba1994JCC,PE:Cic1986CPR,PE:Per1985MM,PE:Ber1984BOOK,PE:Ber1983BOOK,PE:Pea1979JCP,PE:Cha1979JCP,PE:Hel1979JCP,PE:Fix1978JCP}.
In our opinion, the question whether the rigid or the stiff model
should be used to approximate the real quantum mechanical statistics
of an arbitrary organic molecule has not been satisfactorily answered
yet. For discussions about the topic, see references
\cite{PE:Mor2004ACP,PE:Alv2002MP,PE:Alv2000MTS,PE:Hes2002JCP,PE:Ral1979JFM,PE:Hel1979JCP,PE:Go1976MM,PE:Go1969JCP}.
In this work, we adopt the cautious position that any of the two
models may be useful in certain cases or for certain purposes and we
study them both on equal footing. Our concern is, then, to study the
effects that either way of imposing constraints causes in the
conformational equilibrium of macromolecules.

In the Born-Oppenheimer approximation \cite{PE:Bor1927APL} customarily
used in Quantum Mechanics and in the majority of the classical force
fields, the relevant degrees of freedom are the Euclidean (also called
\emph{Cartesian} by some authors) $3n$ coordinates of the $n$
nuclei. However, it is frequent to define a different set of
coordinates in which the overall translation and rotation of the
system are distinguished and the remaining $3n-6$ degrees of freedom
are chosen (according to different prescriptions as \emph{internal
coordinates}, which are simple geometrical parameters (typically
consisting of bond lengths, bond angles and dihedral angles) that
describe the internal structure of the system \cite{PE:Wil1980BOOK}.

In macromolecules, the natural constraints are those derived from the
relative rigidity of the internal covalent structure of groups of
atoms that share a common center (and also from the rigidity of
rotation around double or triple bonds) compared to the energetically
``cheaper'' rotation around single bonds. In internal coordinates,
these chemical constraints may be directly implemented by asking that
some conveniently selected \emph{hard} coordinates (normally, bond
lengths, bond angles and some dihedrals) have constant values or
values that depend on the remaining \emph{soft} coordinates (see
ref.~\cite{PE:Go1976MM} for a definition). In Euclidean coordinates,
on the other hand, the expression of the constraints is more
cumbersome and complicated procedures
\cite{PE:Zho2000JCP,PE:Bar1995JCC,PE:Cic1986CPR,PE:Ber1983BOOK,PE:And1983JCOP,PE:Ryc1977JCOP}
must be used at each timestep to implement them in Molecular Dynamics
simulations. This is why, in the classical stiff model, as well as in
the rigid one, it is common to use internal coordinates and they are
also the choice throughout this work.

In the equilibrium Statistical Mechanics of both the stiff and rigid
models, the marginal probability density in the coordinate part of the
phase space in these internal coordinates is not proportional to the
naive $\exp [-\beta V_{\Sigma}(q^{i})]$, where $V_{\Sigma}(q^{i})$
denotes the potential energy on the constrained
hypersurface\footnote{\label{foot:notation_later}By $q^{i}$, we denote
the soft internal coordinates of the system. See sec.~\ref{sec:theory}
and the Appendix for a precise definition.}. Instead, some correcting
terms that come from different sources must be added to the potential
energy $V_{\Sigma}(q^{i})$
\cite{PE:Mor2004ACP,PE:Sch2003MP,PE:Den2000MP,PE:Den1998JCP,PE:VGu1989BOOK,PE:Ral1979JFM,PE:Go1976MM}.
These terms involve determinants of mass-metric tensors and also of
the Hessian matrix of the constraining part of the potential (see
sec.~\ref{sec:theory}). If Monte Carlo simulations in the coordinate
space are to be performed
\cite{PE:Din2000JCC,PE:Sch1998JCP,PE:Aba1994JCC,PE:Per1994JCC,PE:Kna1993JCC,PE:Alm1990MP}
and the probability densities that correspond to any of these two
models sampled, the corrections should be included or, otherwise,
showed to be negligible.

Additionally, the three different correcting terms are involved in the
definition of the so-called Fixman's compensating potential
\cite{PE:Fix1974PNAS}, which is frequently used to reproduce the stiff
equilibrium distribution using rigid Molecular Dynamics simulations
\cite{PE:Sch2003MP,PE:Pas2002JCP,PE:Rei2000PD,PE:Den2000MP,PE:Den1998JCP,PE:Per1985MM,PE:Pea1979JCP,PE:Cha1979JCP,PE:Hel1979JCP,PE:Fix1978JCP}.

Customarily in the literature, some of these corrections to the
potential energy are assumed to be independent of the conformation and
thus dropped from the basic expressions. Also, subtly entangled to the
assumptions underlying many classical results, a second type of
approximation is made that consists of assuming that the equilibrium
values of the hard coordinates do not depend on the soft coordinates.

In this work, we measure the conformational dependence of \emph{all
correcting terms} and of the Fixman's compensating potential in the
model dipeptide HCO-{\small L}-Ala-NH$_2$ \emph{without any simplifying
assumption}. The potential energy function is considered to be the
effective Born-Oppenheimer potential for the nuclei derived from ab
initio quantum mechanical calculations including electron correlation
at the MP2 level. We also repeat the calculations, with the same basis
set (6-31++G(d,p)) and at the Hartree-Fock level of the theory in
order to investigate if this less demanding method without electron
correlation may be used in further studies. It is \emph{the first
time}, as far as we are aware, that this type of study is performed in
a relevant biomolecule with a realistic potential energy function.

In sec.~\ref{sec:theory}, we introduce the notation to be used and
derive the Statistical Mechanics formulae of the rigid and stiff
models in the general case. In sec.~\ref{sec:methods}, we describe the
computational methods used and we summarize the factorization of the
external coordinates presented in
ref.~\cite{PE:Ech2006JCCc}. Sec.~\ref{sec:results} is devoted to the
presentation and discussion of the assessment of the approximation
that consists of neglecting the different corrections to the potential
energy in the model dipeptide HCO-{\small L}-Ala-NH$_2$, without any
simplifying assumption, which is the central aim of this work. The
conclusions are summarized in sec.~\ref{sec:conclusions}. Finally, in
the appendix, we discuss the use of the different approximations in
the literature and we give a precise definition of \emph{exactly} and
\emph{approximately separable hard and soft coordinates} which will
shed some light on the relation between the different types of
simplifications aforementioned.

\section{Theory}
\label{sec:theory}

First of all, it is convenient to introduce certain notational
conventions that will be used extensively in the rest of the work:

\begin{itemize}

\item The system under scrutiny will be a set of $n$ mass points
 termed \emph{atoms}. The Euclidean coordinates of the atom $\alpha$
 in a set of axes fixed in space are denoted by $\vec{x}_\alpha$. The
 subscript $\alpha$ runs from 1 to $n$.

\item The curvilinear coordinates suitable to describe the system will
 be denoted by $q^{\mu},\ \mu=1,\ldots,3n$ and the set of Euclidean
 coordinates by $x^{\mu}$ when no explicit reference to the atoms
 index needs to be made. We shall often use $N:=3n$ for the total
 number of degrees of freedom.

\item The coordinates $q^{\mu}$ are split into $(q^{A},q^{a}),\
 a=7,\dots,N$. The first six are termed \emph{external coordinates}
 and are denoted by $q^{A}$. They describe the overall position and
 orientation of the system with respect to a frame fixed in space (see
 ref.~\citen{PE:Ech2006JCCc} for further details). The coordinates
 $q^{a}$ are said \emph{internal coordinates} and determine the
 positions of the atoms in the frame fixed in the system. They
 parameterize what we shall call the \emph{internal subspace} or
 \emph{conformational space}, denoted by $\mathcal{I}$ and the
 coordinates $q^{A}$ parameterize the \emph{external subspace},
 denoted by $\mathcal{E}$.

\item The \emph{general set-up} of the problem may be described as
 follows: Instead of us being interested on the conformational
 equilibrium of the system in the external subspace $\mathcal{E}$ plus
 the whole internal subspace $\mathcal{I}$ (i.e., the \emph{whole
 space}, denoted by $\mathcal{E} \times \mathcal{I}$), we wish to find
 the probability density on a hypersurface $\Sigma \subset
 \mathcal{I}$ of dimension $M$ (plus the external subspace
 $\mathcal{E}$), i.e., on $\mathcal{E} \times \Sigma$.

\item In typical internal coordinates $q^{a}$, normally consisting of
 bond lengths, bond angles and dihedral angles (see
 ref.~\citen{PE:Ech2006JCCa} and references therein), the hypersurface
 $\Sigma$ is described via $L:=N-M-6$ constraints:

\begin{equation}
\label{eq:real_constraints}
q^{I}=f^{I}(q^{i}) \qquad I=M+7,\ldots,N \ ,
\end{equation}

 where the $q^{a}$ are split into $q^{a} \equiv (q^{i},q^{I})$, and
 the $q^{i}$, $i=7,\ldots,M+6$, which parameterize $\Sigma$, are
 called \emph{internal soft coordinates}, whereas the $q^{I}$ are
 termed \emph{hard coordinates}. The external coordinates $q^{A}$,
 together with the $q^{i}$, form the whole set of \emph{soft
 coordinates}, denoted by $q^{u} \equiv (q^{A},q^{i})$,
 $u=1,\ldots,M+6$.

\end{itemize}

In table~\ref{tab:def_indices}, a summary of the indices used is
given.

\begin{table}[!ht]
\begin{center}
\begin{tabular}{llll}
Indices & Range & Number & Description \\
\hline \\[-8pt]
$\alpha,\beta,\gamma,\ldots$ & $1,\ldots,n$ & $n$ &Atoms \\
$\mu,\nu,\rho,\ldots$ & $1,\ldots,N$ & $N=3n$ & All coordinates \\
$A,B,C,\ldots$ & $1,\ldots,6$ & 6 & External coordinates \\
$a,b,c,\ldots$ & $7,\ldots,N$ & $N-6$ & Internal coordinates \\
$i,j,k,\ldots$ & $7,\ldots,M+6$ & $M$ & Soft internal coordinates \\
$I,J,K,\ldots$ & $M+7,\ldots,N$ & $L=N-M-6$ & Hard internal coordinates \\
$u,v,w,\ldots$ & $1,\ldots,M+6$ & $M+6$ & All soft coordinates
\end{tabular}
\end{center}
\caption{\label{tab:def_indices}{\small Definition of the indices used.}}
\end{table}

\subsection{Classical stiff model}
\label{subsec:stiff}

In the classical stiff model, the constraints in
eq.~\ref{eq:real_constraints} are implemented by imposing an strong
energy penalization when the internal conformation of the system,
described by $q^{a}$, departs from the constrained hypersurface
$\Sigma$. To ensure this, we must have that the potential energy
function in $\mathcal{I}$ satisfies certain conditions. First, we
write the potential $V(q^{a})$ as
follows\footnote{\label{foot:add_and_subtract}Note that we have simply
added and subtracted from the total potential energy $V(q^{i},q^{I})
\equiv V(q^{a})$ of the system the same quantity,
$V\big(q^{i},f^{I}(q^{i})\big)$.}:

\begin{equation}
\label{eq:VC}
V(q^{i},q^{I})=\underbrace{V\big(q^{i},f^{I}(q^{i})\big)}
                _{\displaystyle V_{\Sigma}(q^{i})} +
  \underbrace{\Big[V(q^{i},q^{I}) - V\big(q^{i},f^{I}(q^{i})\big)\Big]}
  _{\displaystyle V_{\mathrm{c}}(q^{i},q^{I})} \ .
\end{equation}

Next, we impose the following conditions on the \emph{constraining
potential} \, $V_{\mathrm{c}}(q^{i},q^{I})$ defined above:

\begin{enumerate}
\item[(i)] That $V_{\mathrm{c}}\big(q^{i},f^{I}(q^{i})\big) \le
 V_{\mathrm{c}}(q^{i},q^{I}) \quad \forall q^{i},q^{I}$, i.e., that
 $\Sigma$ be the global minimum of $V_{\mathrm{c}}$ (and, henceforth,
 a local one too) with respect to variations of the hard coordinates.
\item[(ii)] That, for small variations $\Delta q^{I}$ on the hard
 coordinates (i.e., for changes $\Delta q^{I}$ considered as
 physically irrelevant), the associated changes in
 $V_{\mathrm{c}}(q^{i},q^{I})$ are much larger than the thermal energy
 $RT$.
\end{enumerate}

The advantages of this formulation, much similar to that on
\cite{PE:Go1976MM}, are many. First, it sets a convenient framework
for the derivation of the Statistical Mechanics formulae of the
classical stiff model relating it to the fully flexible model in the
whole space $\mathcal{E} \times \mathcal{I}$. Second, it clearly
separates the potential energy on $\Sigma$ from the part that is
responsible of implementing the constraints. Third, contrarily to the
formulation based on delta functions \cite{PE:Sch2003MP}, it allows to
clearly understand the necessity of including the correcting term
associated to the determinant of the Hessian of $V_{\mathrm{c}}$ (see
the derivation that follows). Finally, and more importantly for us, it
provides a direct prescription for calculating $V_{\Sigma}(q^{i})$ and
$\Sigma$ (the Potential Energy Surface (PES), frequently used in
Quantum Chemistry calculations
\cite{PE:Lan2005PSFB,PE:Per2003JCC,PE:Var2002JPCA,PE:Yu2001JMS,PE:Csa1999PBMB})
via geometry optimization at fixed values of the soft coordinates.

We also remark that, in order to satisfy point (ii) above and to allow
the derivation of the different correcting terms that follows and the
validity of the final expressions, the hard coordinates $q^{I}$ must
be indeed hard, however, the \emph{soft coordinates} $q^{i}$ do not
have to be soft (in the sense that they produce energetic changes much
smaller than $RT$ when varied). They may be interesting for some other
reason and hence voluntarily picked to describe the system studied,
\emph{without altering the formulae presented in this section.}
Despite this qualifications, the terms \emph{soft} and \emph{hard}
will be kept in this work for consistence with most of the existing
literature
\cite{PE:All2005BOOK,PE:Mor2004ACP,PE:Fre2002BOOK,PE:VGu1989BOOK,PE:Go1976MM},
although, in some cases, the labels \emph{important} and
\emph{unimportant} (for $q^{i}$ and $q^{I}$ respectively), proposed by
Karplus and Kushick \cite{PE:Kar1981MM}, may be more appropriate.

In the case of the model dipeptide HCO-{\small L}-Ala-NH$_2$
investigated in this work, for example, the barriers in the
Ramachandran angles $\phi$ and $\psi$ may be as large as \mbox{$\sim$
40 $RT$}, however, the study of small dipeptides is normally aimed to
the design of effective potentials for polypeptides
\cite{PE:Mac2004JCC,PE:Bor2003JPCB,PE:Bea1997JACS}, where long-range
interactions in the sequence may compensate these local energy
penalizations. This and the fact that the Ramachandran angles are the
relevant degrees of freedom to describe the conformation of the
backbone of these systems, make it convenient to choose them as
\emph{soft coordinates} $q^{i}$ despite the fact that they may be
energetically hard in the case of the dipeptide HCO-{\small
L}-Ala-NH$_2$. As remarked above, this does not affect the
calculations.

Now, due to condition (ii) above, the statistical weights of the
conformations which lie far away from the constrained hypersurface
$\Sigma$ are negligible and, therefore, it suffices to describe the
system in the vicinity of the equilibrium values of the $q^{I}$. In
this region, for each value of the internal soft coordinates $q^{i}$,
we may expand $V_{\mathrm{c}}(q^{i},q^{I})$ in eq.~\ref{eq:VC} up to
second order in the hard coordinates around $\Sigma$ (i.e., around
\mbox{$q^{I}=f^{I}(q^{i})$}) and drop the higher order terms:

\begin{eqnarray}
\label{eq:VC_Taylor}
V_{\mathrm{c}}(q^{i},q^{I}) & \simeq &
  V_{\mathrm{c}}\big(q^{i},f^{I}(q^{i})\big) +
 \left [ \frac{\partial V_{\mathrm{c}}}{\partial q^{J}} \right ]_{\Sigma}
 \big(q^{J}-f^{J}(q^{i})\big) + \nonumber \\
&& + \frac{1}{2} \underbrace{ \left[ \frac{\partial^{2} 
  V_{\mathrm{c}}}{\partial q^{J}
   \partial q^{K}} \right ]_{\Sigma} }
  _{\displaystyle \mathcal{H}_{JK}(q^{i})} \big(q^{J}-f^{J}(q^{i})\big)
  \big(q^{K}-f^{K}(q^{i})\big) \ ,
\end{eqnarray}

where the subindex $\Sigma$ indicates evaluation on the constrained
hypersurface and a more compact notation, $\mathcal{H}(q^{i})$, has
been introduced for the Hessian matrix of $V_{\mathrm{c}}$ with
respect to the hard variables evaluated on $\Sigma$.  Also, the
Einstein's sum convention is assumed on repeated indices.

In this expression, the zeroth order term
$V_{\mathrm{c}}\big(q^{i},f^{I}(q^{i})\big)$ is zero by definition of
$V_{\mathrm{c}}$ (see eq.~\ref{eq:VC}) and the linear term is also
zero, because of the condition (i) above. Hence, the first non-zero
term of the expansion in eq.~\ref{eq:VC_Taylor} is the second order
one. Using this, together with eq.~\ref{eq:VC}, we may write the
\emph{stiff Hamiltonian}

\begin{eqnarray}
\label{eq:Hs}
H_{\mathrm{s}}(q^{\mu},p_{\mu}) &:=&
    \frac{1}{2}p_{\nu}G^{\nu\rho}(q^{u},q^{I})p_{\rho} +
    V_{\Sigma}(q^{i}) + \nonumber \\
    && +\ \frac{k}{2} \mathcal{H}_{JK}(q^{i})
   \big(q^{J}-f^{J}(q^{i})\big)
   \big(q^{K}-f^{K}(q^{i})\big) \ ,
\end{eqnarray}

the \emph{mass-metric tensor} $G_{\nu\rho}$ being

\begin{equation}
\label{eq:G}
G_{\nu\rho}(q^{u},q^{I}):=\sum_{\sigma=1}^{N}
                \frac{\partial x^{\sigma}(q^{\mu})}
                {\partial q^{\nu}}m_{\sigma}
                \frac{\partial x^{\sigma}(q^{\mu})}{\partial q^{\rho}}
\end{equation}

and $G^{\nu\rho}$ its inverse, defined by

\begin{equation}
\label{eq:inverse}
G^{\nu\sigma}(q^{u},q^{I})\,
 G_{\sigma\rho}(q^{u},q^{I}) = \delta^{\nu}_{\rho} \ ,
\end{equation}

where $\delta^{\nu}_{\rho}$ denotes the Kronecker's delta.

Therefore, the \emph{stiff partition function} of the system
is\footnote{\label{foot:canonical}No Jacobian appears in the integral
measure because $q^{\mu}$ and $p_{\mu}$ are obtained from the
Euclidean coordinates via a canonical transformation
\cite{PE:Arn1989BOOK}.}

\begin{equation}
\label{eq:Zs1}
Z_{\mathrm{s}} = \frac{\alpha_{QM}}{h^{N}} \int \mathrm{d}q^{\mu} 
    \mathrm{d}p_{\mu} \exp \big[ - \beta H_{\mathrm{s}}(q^{\mu},p_{\mu})
     \big ] \ ,
\end{equation}

where $h$ is Planck's constant, we denote $\beta:=1/RT$ (per mole
energy units are used throughout the article, so $RT$ is preferred
over $k_{B}T$) and $\alpha_{QM}$ is a combinatorial number that
accounts for quantum indistinguishability and that must be specified
in each particular case (e.g., for a gas of $N$ indistinguishable
particles, $\alpha_{QM}=1/N!$).

Now, using the condition (ii) again, the $q^{I}$ appearing in the
mass-metric tensor $G$ in $H_{\mathrm{s}}$ (in eq.~\ref{eq:Zs1}) can
be approximately evaluated at their equilibrium values $f^{I}(q^{i})$,
yielding, for the stiff partition function,

\begin{equation}
\label{eq:Zs2}
\begin{array}{l}
\displaystyle Z_{\mathrm{s}} = \frac{\alpha_{QM}}{h^{N}}
    \int \mathrm{d}q^{u} \mathrm{d}q^{I} 
    \mathrm{d}p_{\mu} \exp \bigg[ - \beta \bigg(
    \frac{1}{2}p_{\nu}G^{\nu\rho}
                    \big(q^{u},f^{I}(q^{i})\big)p_{\rho} + \\[10pt]
\displaystyle \quad + \ V_{\Sigma}(q^{i}) +
    \frac{1}{2} \mathcal{H}_{JK}(q^{i})\big({q}^{J}-f^{J}(q^{i})\big)
    \big({q}^{K}-f^{K}(q^{i})\big) \bigg) \bigg] \ .
\end{array}
\end{equation}

If we now integrate over the hard coordinates $q^{I}$, we have

\begin{equation}
\label{eq:Zs3}
\begin{array}{l}
\displaystyle Z_{\mathrm{s}} = 
    \left( \frac{2\pi}{\beta} \right )^{\frac{L}{2}} \frac{\alpha_{QM}}{h^{N}}
    \int \mathrm{d}q^{u} \mathrm{d}p_{\mu} \exp \bigg[ - \beta \bigg(
    \frac{1}{2}p_{\nu}G^{\nu\rho}
                    \big(q^{u},f^{I}(q^{i})\big)p_{\rho} + \\[10pt]
\displaystyle \quad + \ V_{\Sigma}(q^{i}) +
    T \frac{R}{2} \ln \Big[ \mathrm{det}\, \mathcal{H}(q^{i}) \Big]
    \bigg) \bigg] \ .
\end{array}
\end{equation}

where the part of the result of the Gaussian integral consisting of
${\det}^{-1/2}\mathcal{H}$ has been taken to the exponent.

Note that the Hessian matrix $\mathcal{H}_{JK}$ involves only
derivatives with respect to the hard coordinates (see
eq.~\ref{eq:VC_Taylor}), so that the minimization protocol embodied in
eq.~\ref{eq:VC} (which is identical to the procedure followed in
Quantum Chemistry for computing the PES along reaction coordinates)
guarantees that $\mathcal{H}_{JK}$ is positive defined and, hence,
${\det}\,\mathcal{H}$ is positive, allowing to take its logarithm as
in the previous expression. The fact that it is only this `partial
Hessian' that makes sense in the computation of equilibrium properties
along soft (or reaction) coordinates, has been recently pointed out in
ref.~\citen{PE:Vis2006JPCA}.

It is also frequent to integrate over the momenta in the partition
function. Doing this in eq.~\ref{eq:Zs3} and taking the determinant
of the mass-metric tensor that shows
up\footnote{\label{foot:determinant}Note that, by $G$, we denote the
matrix that corresponds to the mass-metric tensor with two covariant
indices $G_{\mu\nu}$. The same convention has been followed for the
Hessian matrix $\mathcal{H}$ in eq.~\ref{eq:Zs3} and for the reduced
mass-metric tensor $g$ in eq.~\ref{eq:Zr2}.} to the exponent, we may
write the partition function as an integral only on the coordinates:

\begin{equation}
\label{eq:Zs4}
\begin{array}{l}
\displaystyle Z_{\mathrm{s}}=\chi_{\mathrm{s}}(T) \int \mathrm{d}q^{u}
  \exp \bigg[ - \beta \bigg( V_{\Sigma}(q^{i}) +\ T\frac{R}{2}
  \ln\Big[\mathrm{det}\,\mathcal{H}(q^{i})\Big] - \\[10pt]
  \displaystyle \quad - T\frac{R}{2}
  \ln\Big[\mathrm{det}\,G\big(q^{u},f^{I}(q^{i})\big)\Big]
  \bigg) \bigg] \ ,
\end{array}
\end{equation}

where the multiplicative factor that depends on $T$ has been
defined as follows:

\begin{equation}
\label{eq:chiTs}
\chi_{\mathrm{s}}(T):=\left (\frac{2\pi}{\beta}\right )^{\frac{N+L}{2}}
 \frac{\alpha_{QM}}{h^{N}} \ .
\end{equation}

If the exponent in eq.~\ref{eq:Zs4} is seen as a free energy, then,
$V_{\Sigma}(q^{i})$ may be regarded as the internal energy and the two
conformation-dependent correcting terms that are added to it as
effective entropies (which is compatible with their being linear in
$RT$). The second one comes only from the desire to write the marginal
probabilities in the coordinate space (i.e., averaging the momenta)
and may be called a \emph{kinetic entropy} \cite{PE:Go1969JCP}, the
first term, on the other hand, is truly an entropic term that comes
from the averaging out of certain degrees of freedom and it is
reminiscent of the \emph{conformational} or \emph{configurational
entropies} appearing in quasiharmonic analysis
\cite{PE:And2001JCP,PE:Kar1981MM,PE:VGu1982MM}.

In this spirit, we define

\begin{subequations}
\label{eq:Ss}
\begin{align}
& F_{\mathrm{s}}(q^{u}):= V_{\Sigma}(q^{i}) -
   T \big( S_{\mathrm{s}}^{c}(q^{i}) + 
   S_{\mathrm{s}}^{\mathrm{k}}(q^{u}) \big)\  ,
\label{eq:Fs} \\
& S_{\mathrm{s}}^{c}(q^{i}):=-\frac{R}{2}\ln\Big[\mathrm{det}\,
                   \mathcal{H}(q^{i})\Big] \  ,
\label{eq:Ssc} \\
& S_{\mathrm{s}}^{\mathrm{k}}(q^{u}):=\frac{R}{2}\ln\Big[\mathrm{det}\,
                   G\big(q^{u},f^{I}(q^{i})\big)\Big] \  .
\label{eq:Ssk}
\end{align}
\end{subequations}

In such a way that the \emph{stiff equilibrium probability} in the soft
subspace $\mathcal{E} \times \Sigma$ is given by

\begin{equation}
\label{eq:Ps}
P_{\mathrm{s}}(q^{u}) = \frac{\exp \big[-\beta F_{\mathrm{s}}(q^{u})\big]}
                             {\displaystyle Z^{\,\prime}_{\mathrm{s}}} \ ,
\quad \mathrm{with} \quad
Z^{\,\prime}_{\mathrm{s}} := \int \mathrm{d}q^{u}
                             \exp \big[-\beta F_{\mathrm{s}}(q^{u})\big] \ .
\end{equation}

Now, it is worth remarking that, although the kinetic entropy
$S_{\mathrm{s}}^{\mathrm{k}}$ depends on the external coordinates
$q^{A}$, we have recently shown \cite{PE:Ech2006JCCc} that the
determinant of the mass-metric tensor $G$ may be written, for any
molecule, general internal coordinates and arbitrary constraints, as a
product of two functions; one depending only on the external
coordinates, and the other only on the internal ones $q^{a}$. Hence
the externals-dependent factor in eq.~\ref{eq:Ssk} may be integrated
out independently to yield an effective free energy and a probability
density $P_{\mathrm{s}}$ that depend only on the soft internals
$q^{i}$ (see sec.~\ref{subsec:externals}).

\subsection{Classical rigid model}
\label{subsec:rigid}

If the relations in eq.~\ref{eq:real_constraints} are considered to
hold \emph{exactly} and are treated as holonomic constraints, the
Hamiltonian function that describes the Classical Mechanics in the
subspace $(\mathcal{E} \times \Sigma) \subset (\mathcal{E} \times
\mathcal{I})$, spanned by the coordinates $q^{u}$, may be written as
follows:

\begin{equation}
\label{eq:Hr}
H_{\mathrm{r}}(q^{u},\eta_{u}):=
 \frac{1}{2}\eta_{v}g^{vw}(q^{u})\eta_{w} + V_{\Sigma}(q^{i}) \ ,
\end{equation}

where the \emph{reduced mass-metric tensor} $g_{vw}(q^{u})$ in
$\mathcal{E} \times \Sigma$, that appears in the kinetic energy, is
(see what follows)

\begin{equation}
\label{eq:g}
\begin{array}{l}
\displaystyle g_{vw}(q^{u}) = G_{vw}\big(q^{u},f^{I}(q^{i})\big) +
 \frac{\partial f^{J}(q^{i})}{\partial q^{v}}
 G_{JK}\big(q^{u},f^{I}(q^{i})\big)
 \frac{\partial f^{K}(q^{i})}{\partial q^{w}}\ + \\[10pt]
\displaystyle \quad + \ G_{vK}\big(q^{u},f^{I}(q^{i})\big)
 \frac{\partial f^{K}(q^{i})}{\partial q^{w}} +
 \frac{\partial f^{J}(q^{i})}{\partial q^{v}}
 G_{Jw}\big(q^{u},f^{I}(q^{i})\big) := \\[10pt]
\displaystyle \quad = \,
 \frac{\partial \tilde{f}^{\mu}}{\partial q^{v}}
 G_{\mu\nu}\big(q^{u},f^{I}(q^{i})\big)
 \frac{\partial \tilde{f}^{\nu}}{\partial q^{w}} \ ,
\end{array}
\end{equation}

and $g^{vw}(q^{u})$ is defined to be its inverse in the sense of
eq.~\ref{eq:inverse}. Also, the notation

\begin{equation}
\label{eq:ftilde}
\tilde{f}^{\mu}:=
\begin{cases}
q^{u} & \quad \mathrm{if} \quad u:=\mu = 1,\ldots,M+6 \\
f^{I}(q^{i}) & \quad \mathrm{if} \quad I:=\mu = M+7,\ldots,N \\
\end{cases}
\end{equation}

has been introduced for convenience.

Note that eq.~\ref{eq:g} may derived from the unconstrained
Hamiltonian in $(\mathcal{E} \times \mathcal{I})$,

\begin{equation}
\label{eq:H}
H(q^{\mu},p_{\mu}):=\frac{1}{2}p_{\nu}G^{\nu\rho}(q^{\mu})p_{\rho}+V(q^{a}) \ ,
\end{equation}

using the constraints in eq.~\ref{eq:real_constraints}, together with
its time derivatives (denoted by an overdot: as in $\dot{A}$)

\begin{equation}
\label{eq:diff_constraints}
 \dot{q}^{I}:=\frac{\partial f^{I}(q^{i})}{\partial q^{j}}
 \dot{q}^{j}
\end{equation}

and defining the momenta $\eta_{v}$ as

\begin{equation}
\label{eq:eta}
 \eta_{v}:= g_{vw}(q^{u})\,\dot{q}^{w} = g_{vw}(q^{u})\, 
 G^{w\mu}\big(q^{u},f^{I}(q^{i})\big)\,p_{\mu} \ .
\end{equation}

Hence, the \emph{rigid partition function} is

\begin{equation}
\label{eq:Zr1}
Z_{\mathrm{r}} = \frac{\alpha_{QM}}{h^{M+6}} \int \mathrm{d}q^{u} 
    \mathrm{d}\eta_{u} \exp{\left [ - \beta \left (
    \frac{1}{2}\eta_{v}g^{vw}(q^{u})\eta_{v} +
    V_{\Sigma}(q^{i}) \right ) \right ] } \ .
\end{equation}

Integrating over the momenta, we obtain the marginal probability
density in the coordinate space analogous to eq.~\ref{eq:Zs4}:

\begin{equation}
\label{eq:Zr2}
Z_{\mathrm{r}} = \chi_{\mathrm{r}}(T) \int \mathrm{d}q^{u}
    \exp{\Big[ - \beta \Big( V_{\Sigma}(q^{i}) - 
    T \frac{R}{2} \ln \Big[ \mathrm{det}\,g(q^{u}) 
    \Big] \Big) \Big]} \ ,
\end{equation}

where

\begin{equation}
\label{eq:chiTr}
\chi_{\mathrm{r}}(T):=\left (\frac{2\pi}{\beta}\right )^{\frac{M+6}{2}}
 \frac{\alpha_{QM}}{h^{\frac{M+6}{2}}} \ .
\end{equation}

Repeating the analogy with free energies and entropies in the last
paragraphs of the previous subsection, we define

\begin{subequations}
\label{eq:Sr}
\begin{align}
& F_{\mathrm{r}}(q^{u}):= V_{\Sigma}(q^{i}) -
   T S_{\mathrm{r}}^{\mathrm{k}}(q^{u}) \  ,
\label{eq:Fr} \\
& S_{\mathrm{r}}^{\mathrm{k}}(q^{u}):=
  \frac{R}{2}\ln\Big[\mathrm{det}\,g(q^{u})\Big] \ ,
\label{eq:Srk}
\end{align}
\end{subequations}

being the \emph{rigid equilibrium probability} in the soft
subspace $\mathcal{E} \times \Sigma$

\begin{equation}
\label{eq:Pr}
P_{\mathrm{r}}(q^{u}) = \frac{\exp \big[-\beta F_{\mathrm{r}}(q^{u})\big]}
                             {\displaystyle Z^{\,\prime}_{\mathrm{r}}} \ ,
\quad \mathrm{with} \quad
Z^{\,\prime}_{\mathrm{r}} := \int \mathrm{d}q^{u}
                             \exp \big[-\beta F_{\mathrm{r}}(q^{u})\big] \ .
\end{equation}

As in the case of $G$, we have shown in ref.~\cite{PE:Ech2006JCCc} that
the determinant of the reduced mass-metric tensor $g$ may be written,
for any molecule, general internal coordinates and arbitrary
constraints, as a product of two functions; one depending only on the
external coordinates, and the other only on the internal ones $q^{i}$.
Hence the externals-dependent factor in $\mathrm{det}\,g(q^{u})$ may
be integrated out independently to yield a free energy and a
probability density $P_{\mathrm{r}}$ that depend only on the soft
internals $q^{i}$ (see sec~\ref{subsec:externals}).

To end this subsection, we remark that it is frequent in the
literature
\cite{PE:Mor2004ACP,PE:Sch2003MP,PE:Pas2002JCP,PE:Rei2000PD,PE:Den2000MP,PE:Alm1990MP,PE:Per1985MM,PE:Pea1979JCP,PE:Cha1979JCP,PE:Fix1978JCP}
to define the so-called \emph{Fixman's compensating potential}
\cite{PE:Fix1974PNAS} as the difference between
$F_{\mathrm{s}}(q^{u})$, in eq.~\ref{eq:Ss}, and
$F_{\mathrm{r}}(q^{u})$, defined above, i.e.,

\begin{eqnarray}
\label{eq:VF}
V_{\mathrm{F}}(q^{u}) & := & T S_{\mathrm{r}}^{\mathrm{k}}(q^{u}) -
 T S_{\mathrm{s}}^{c}(q^{i}) - 
 T S_{\mathrm{s}}^{\mathrm{k}}(q^{u}) = \nonumber \\
& = &  \frac{RT}{2}\ln\Bigg [\frac{\mathrm{det}\,G(q^{u})}{\mathrm{det}\,
     \mathcal{H}(q^{i})\, \mathrm{det}\,g(q^{u})}\Bigg] \ .
\end{eqnarray}

Hence, performing rigid Molecular Dynamics simulations, which would
yield an equilibrium distribution proportional to $\exp [-\beta
F_{\mathrm{r}}(q^{u})]$, and adding $V_{\mathrm{F}}(q^{u})$ to the
potential energy $V_{\Sigma}(q^{i})$ one can reproduce instead the
stiff probability density $P_{\mathrm{s}} \propto \exp [-\beta
F_{\mathrm{s}}(q^{u})]$
\cite{PE:Mor2004ACP,PE:Sch2003MP,PE:Pas2002JCP,PE:Den2000MP,PE:Den1998JCP,PE:Per1985MM,PE:Pea1979JCP,PE:Cha1979JCP,PE:Hel1979JCP,PE:Fix1978JCP}.
This allows to obtain at a lower computational cost (due to the
timescale problems discussed in the introduction) equilibrium averages
that otherwise must be extracted from expensive fully flexible
whole-space simulations. In fact, it seems that this particular
application of the theoretical tools herein described, and not the
search for the correct probability density to sample in Monte Carlo
simulations, was what prompted the interest in the study of
mass-metric tensors effects.

Finally, in table~\ref{tab:correcting_terms} we summarize the
equilibrium probability densities and the different correcting terms
derived in this section.

\begin{table}[!ht]
\begin{equation}
\begin{array}{l@{\hspace{20pt}}l}
\mathrm{Classical\ Stiff\ Model} & \mathrm{Classical\ Rigid\ Model} \\
\hline\\[-8pt]
P_{\mathrm{s}}(q^{u}) = \frac{\displaystyle \exp
                              \big[-\beta F_{\mathrm{s}}(q^{u})\big]}
                             {\displaystyle Z^{\,\prime}_{\mathrm{s}}} & 
P_{\mathrm{r}}(q^{u}) = \frac{\displaystyle \exp
                              \big[-\beta F_{\mathrm{r}}(q^{u})\big]}
                             {\displaystyle Z^{\,\prime}_{\mathrm{r}}} \\[8pt]
F_{\mathrm{s}}(q^{u}):= V_{\Sigma}(q^{i}) -
   T \big( S_{\mathrm{s}}^{c}(q^{i}) + 
   S_{\mathrm{s}}^{\mathrm{k}}(q^{u}) \big) & 
F_{\mathrm{r}}(q^{u}):= V_{\Sigma}(q^{i}) -
   T S_{\mathrm{r}}^{\mathrm{k}}(q^{u}) \\[6pt]
\displaystyle S_{\mathrm{s}}^{\mathrm{k}}(q^{u}):=
                   \frac{R}{2}\ln\Big[\mathrm{det}\,
                   G\big(q^{u},f^{I}(q^{i})\big)\Big] & 
\displaystyle S_{\mathrm{r}}^{\mathrm{k}}(q^{u}):=
  \frac{R}{2}\ln\Big[\mathrm{det}\,g(q^{u})\Big] \\[8pt]
\displaystyle S_{\mathrm{s}}^{c}(q^{i}):=-\frac{R}{2}\ln\Big[\mathrm{det}\,
                   \mathcal{H}(q^{i})\Big] & 
\end{array} \nonumber
\end{equation}
\caption{\label{tab:correcting_terms}{\small Equilibrium probability
densities and correcting terms to the potential energy
$V_{\Sigma}(q^{i})$ in the classical stiff and rigid models of
constraints.}}
\end{table}

\section{Methods}
\label{sec:methods}

\subsection{Factorization of the external coordinates}
\label{subsec:externals}

In the recent work \cite{PE:Ech2006JCCc}, we have shown that
the determinant of the mass-metric tensor $G$ in
eq.~\ref{eq:Ssk} can be written as follows if the SASMIC
\cite{PE:Ech2006JCCa} coordinates for general branched molecules
are used:

\begin{equation}
\label{eq:detG_sasmic}
\det G = \left( \prod_{\alpha=1}^{n} m_{\alpha}^{3} \right)
         {\sin}^{2}\theta \left( \prod_{\alpha=2}^{n} r_{\alpha}^{4} \right)
          \left( \prod_{\alpha=3}^{n} {\sin}^{2}\theta_{\alpha} \right) \ ,
\end{equation}

where the $r_{\alpha}$ are bond lengths and the $\theta_{\alpha}$ bond
angles.

Note that this expression, whose validity was proved for the more
particular case of serial polymers by G\=o and Scheraga
\cite{PE:Go1976MM} and, before, by Volkenstein
\cite{PE:Vol1959BOOKpp}, does not explicitly depend on the dihedral
angles. However, it may depend on them via the hard coordinates if the
constraints in the form presented in eq.~\ref{eq:real_constraints}
are used.

The term depending on the masses of the atoms in the expression above
may be dropped from eq.~\ref{eq:Ssk}, because it does not depend on
the conformation, and the only part of $\det G$ that depend on the
external coordinates, ${\sin}^{2}\theta$, may be integrated out in
eq.~\ref{eq:Zs4} ($\theta$ is one of the externals $q^{A}$ that
describe the overall orientation of the molecule; see
ref.~\citen{PE:Ech2006JCCc} for further details). Hence, the kinetic
entropy due to the mass-metric tensor $G$ in the stiff case, may be
written, up to additive constants, as

\begin{equation}
\label{eq:Ssk2}
S_{\mathrm{s}}^{\mathrm{k}}(q^{i})=\frac{R}{2}\Bigg[\sum_{\alpha=2}^{n}\ln
 \big(r_{\alpha}^{4}\big)
 + \sum_{\alpha=3}^{n}\ln\big({\sin}^{2}\theta_{\alpha}\big)\Bigg] \  ,
\end{equation}

where the individual contributions of each degree of freedom
have been factorized.

Also in reference \cite{PE:Ech2006JCCc}, we have shown that
the determinant of the reduced mass-metric tensor $g$ in
eq.~\ref{eq:Srk} can be written as follows:

\begin{equation}
\label{eq:detg}
\det g={\sin}^{2}\theta \, \det g_{2}(q^{i}) \ ,
\end{equation}

being the matrix $g_{2}$

\begin{equation}
\label{eq:g2}
g_{2}=
\left( \begin{array}{cc|c@{\hspace{2pt}}c@{\hspace{2pt}}c}
m_{tot}\displaystyle I^{(3)} & \displaystyle m_{tot}\,v(\vec{R}) &
\cdots & \displaystyle m_{tot} \frac{\partial \vec{R}}{\partial q^{j}} &
\cdots \\[10pt]
\displaystyle m_{tot}\,v^{T}(\vec{R}) & \mathcal{J} &
\cdots & \displaystyle \mathop{\Sigma}_{\alpha}m_{\alpha}
\frac{\partial \vec{x}^{\,\prime}_{\alpha}}{\partial q^{j}} \times
\vec{x}^{\,\prime}_{\alpha} & \cdots \\[10pt]
\hline
\vdots & \vdots & & \vdots & \\
\displaystyle m_{tot}\frac{\partial \vec{R}}{\partial q^{i}} &
\displaystyle \mathop{\Sigma}_{\alpha}m_{\alpha} \left(
\frac{\partial \vec{x}^{\,\prime}_{\alpha}}{\partial q^{i}} \times
\vec{x}^{\,\prime}_{\alpha} \right)^{T}&
\cdots &
\displaystyle \mathop{\Sigma}_{\alpha}m_{\alpha}
\frac{\partial \vec{x}^{\,\prime\,T}_{\alpha}}{\partial q^{i}}
\frac{\partial \vec{x}^{\,\prime}_{\alpha}}{\partial q^{j}} & \cdots \\[-2pt]
\vdots & \vdots & & \vdots & 
\end{array} \right) \ 
\end{equation}

where the superindex $T$ indicates matrix transposition, $I^{(3)}$
denotes the $3 \times 3$ identity matrix and
$\vec{x}^{\,\prime}_{\alpha}$ is the position of atom $\alpha$ in the
reference frame fixed in the system (the `primed' reference frame).

Additionally, we denote the \emph{total mass} of the system by
$m_{tot} := \sum_{\alpha}m_{\alpha}$, the position of the \emph{center
of mass} of the system in the primed reference frame by $\vec{R} :=
m^{-1}_{tot}\sum_{\alpha}m_{\alpha} \vec{x}^{\,\prime}_{\alpha}$ and
the \emph{inertia tensor} of the system, also in the primed reference
frame, by

\begin{equation}
\label{eq:inertia}
\mathcal{J}\!:=\!\!
\left( \begin{smallmatrix}
\sum_{\alpha}m_{\alpha} ((x^{\,\prime\,2}_{\alpha})^{2}+
                        (x^{\,\prime\,3}_{\alpha})^{2}) &
-\sum_{\alpha}m_{\alpha}x^{\,\prime\,1}_{\alpha}
                        x^{\,\prime\,2}_{\alpha} &
-\sum_{\alpha}m_{\alpha}x^{\,\prime\,1}_{\alpha}
                        x^{\,\prime\,3}_{\alpha} \\
-\sum_{\alpha}m_{\alpha}x^{\,\prime\,1}_{\alpha}
                        x^{\,\prime\,2}_{\alpha} &
\sum_{\alpha}m_{\alpha} ((x^{\,\prime\,1}_{\alpha})^{2}+
                        (x^{\,\prime\,3}_{\alpha})^{2}) &
-\sum_{\alpha}m_{\alpha}x^{\,\prime\,2}_{\alpha}
                        x^{\,\prime\,3}_{\alpha} \\
-\sum_{\alpha}m_{\alpha}x^{\,\prime\,1}_{\alpha}
                        x^{\,\prime\,3}_{\alpha} &
-\sum_{\alpha}m_{\alpha}x^{\,\prime\,2}_{\alpha}
                        x^{\,\prime\,3}_{\alpha} &
\sum_{\alpha}m_{\alpha} ((x^{\,\prime\,1}_{\alpha})^{2}+
                        (x^{\,\prime\,2}_{\alpha})^{2}) \\
\end{smallmatrix} \right)
 .
\end{equation}

The matrix $v(\vec{R})$ is defined as:

\begin{equation}
\label{eq:vR}
v(\vec{R}):=
\left( \begin{array}{ccc}
0 & -R^{3} & R^{2} \\
R^{3} & 0 & -R^{1} \\
-R^{2} & R^{1} & 0
\end{array} \right)
\end{equation}

and $\times$ denotes the usual vector cross product.

Then, since ${\sin}^{2}\theta$ may be integrated out in eq.~\ref{eq:Zr2},
we can write, omitting additive constants, the kinetic entropy
associated to the reduced mass-metric tensor $g$ depending only on the
soft internals $q^{i}$:

\begin{equation}
\label{eq:Srk2}
S_{\mathrm{r}}^{\mathrm{k}}(q^{i})=
  \frac{R}{2}\ln\Big[\mathrm{det}\,g_{2}(q^{i})\Big]\  .
\end{equation}

Finally, one may note that, since ${\sin}^{2}\theta$ divides out in
the second line of eq.~\ref{eq:VF} or, otherwise stated,
eqs.~\ref{eq:Ssk2} and \ref{eq:Srk2} may be introduced in the
first line, then the Fixman's potential is independent of the external
coordinates as well.

\subsection{Computational Methods}
\label{subsec:comput_methods}

\begin{figure}[!ht]
\begin{center}
\epsfig{file=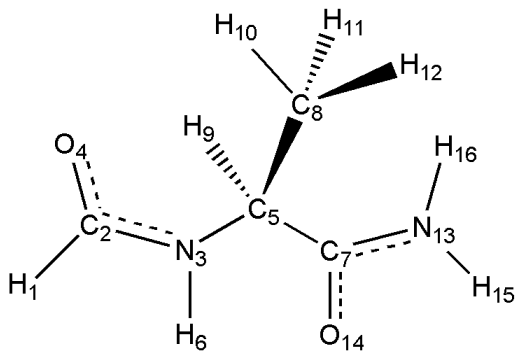,width=7cm}
\end{center}
\caption{\label{fig:num_ala}{\small Atom numeration of the protected
dipeptide HCO-{\small L}-Ala-NH$_2$.}}
\end{figure}

In the particular molecule treated in this work (the model dipeptide
HCO-{\small L}-Ala-NH$_2$ in fig.~\ref{fig:num_ala}), the formulae in
the preceding sections must be used with $M=2$, being the internal
soft coordinates $q^{i} \equiv (\phi,\psi)$ the typical Ramachandran
angles \cite{PE:Ram1963JMB} (see table~\ref{tab:coor_ala}), the total
number of coordinates $N=48$ and the number of hard internals
$L=40$.

Regarding the side chain angle $\chi$, it has been argued elsewhere
\cite{PE:Ech2006JCCa} that it is soft with the same right as the
angles $\phi$ and $\psi$, i.e., the barriers that hinder the rotation
on this dihedral are comparable to the ones existing in the
Ramachandran surface. However, the height of these barriers is
sufficient (\mbox{$\sim$ 6-12 $RT$}, see ref.~\cite{PE:Ech2006JCCa})
for the condition (ii) in sec.~\ref{subsec:stiff} to hold and,
therefore, its inclusion in the set of hard coordinates is convenient
due to its \emph{unimportant} character (see discussion in
sec.~\ref{subsec:stiff}). Moreover, to describe the behaviour
associated to $\chi$ with a probability density different from a
Gaussian distribution (i.e., its potential energy different from an
harmonic oscillator), for example with the tools used in the field of
circular statistics \cite{PE:Hni2003JCC,PE:Dem2001MP,PE:Mar2000BOOKb},
would severely complicate the derivation of the classical stiff model
without adding any conceptual insight to the problem. In addition,
although $\chi$ is a periodic coordinate with threefold symmetry, the
considerable height of the barriers between consecutive minima allows
to make the quadratic assumption in eq.~\ref{eq:VC_Taylor} at each
equivalent valley and permits the approximation of the integral on
$\chi$ by three times a Gaussian integral. The multiplicative factor 3
simply adds a temperature- and conformation-independent reference to
the configurational entropy $S_{\mathrm{s}}^{c}$ in
eq.~\ref{eq:Ssc}.

The same considerations are applied to the dihedral angles,
$\omega_{0}$ and $\omega_{1}$ (see table~\ref{tab:coor_ala}), that
describe the rotation around the peptide bond, and the quadratic
approximation described above can also be used, since the heights of
the rotation barriers around these degrees of freedom are even larger
than the ones in the case of $\chi$.

The ab initio quantum mechanical calculations have been done with the
package GAMESS \cite{PE:Sch1993JCC} under Linux and in 3.20 GHz PIV
machines. The coordinates used for the HCO-{\small L}-Ala-NH$_2$
dipeptide in the GAMESS input files and the ones used to generate them
with automatic Perl scripts are the SASMIC coordinates introduced in
ref.~\citen{PE:Ech2006JCCa}. They are presented in
table~\ref{tab:coor_ala} indicating the name of the conventional
dihedral angles (see also fig.~\ref{fig:num_ala} for reference). To
perform the energy optimizations, however, they have been converted to
Delocalized Coordinates \cite{PE:Bak1996JCP} in order to accelerate
convergence.

\begin{table}[!ht]
\begin{center}
\begin{tabular}{cccc}
Atom name & Bond length & Bond angle & Dihedral angle \\
\hline \\[-8pt]
H$_{1}$ & & & \\
C$_{2}$ & (2,1) & & \\
N$_{3}$ & (3,2) & (3,2,1) & \\
O$_{4}$ & (4,2) & (4,2,1) & (4,2,1,3) \\
C$_{5}$ & (5,3) & (5,3,2) & $\omega_0:=${\bf (5,3,2,1)} \\
H$_{6}$ & (6,3) & (6,3,2) & (6,3,2,5) \\
C$_{7}$ & (7,5) & (7,5,3) & $\phi:=${\bf (7,5,3,2)} \\
C$_{8}$ & (8,5) & (8,5,3) & (8,5,3,7) \\
H$_{9}$ & (9,5) & (9,5,3) & (9,5,3,7) \\
H$_{10}$ & (10,8) & (10,8,5) & $\chi:=${\bf (10,8,5,3)} \\
H$_{11}$ & (11,8) & (11,8,5) & (11,8,5,10) \\
H$_{12}$ & (12,8) & (12,8,5) & (12,8,5,10) \\
N$_{13}$ & (13,7) & (13,7,5) & $\psi:=${\bf (13,7,5,3)} \\
O$_{14}$ & (14,7) & (14,7,5) & (14,7,5,13) \\
H$_{15}$ & (15,13) & (15,13,7) & $\omega_1:=${\bf (15,13,7,5)} \\
H$_{16}$ & (16,13) & (16,13,7) & (16,13,7,15)
\end{tabular}
\end{center}
\caption{\label{tab:coor_ala}{\small SASMIC internal coordinates
(Echenique P. and Alonso J. L., \emph{To be published in
J. Comp. Chem.}, {\tt arXiv:q-bio.BM/0511004}) in Z-matrix form of the
protected dipeptide HCO-{\small L}-Ala-NH$_2$. Principal dihedrals are
indicated in bold face and their typical biochemical name is given.}}
\end{table}

First, we have calculated the typical Potential Energy Surface (PES)
in a regular 12x12 grid of the bidimensional space spanned by the
Ramachandran angles $\phi$ and $\psi$, with both angles ranging from
$-165^{o}$ to $165^{o}$ in steps of $30^{o}$.  This has been done by
running constrained energy optimizations at the
\mbox{MP2/6-31++G(d,p)} level of the theory, freezing the two
Ramachandran angles at each value of the grid, starting from
geometries previously optimized at a lower level of the theory and
setting the gradient convergence criterium to {\tt OPTTOL}=$10^{-5}$
and the self-consistent Hartree-Fock convergence criterium to {\tt
CONV}=$10^{-6}$.

The results of these calculations (which took $\sim$ 100 days of CPU
time) are 144 conformations that define $\Sigma$ and the values of
$V_{\Sigma}(\phi,\psi)$ at these points (the PES itself).

Then, at each optimized point of $\Sigma$, we have calculated the
Hessian matrix in the coordinates of table~\ref{tab:coor_ala} removing
the rows and columns corresponding to the soft angles $\phi$ and
$\psi$, the result being the matrix $\mathcal{H}(\phi,\psi)$ in
eq.~\ref{eq:Ssc}. This has been done, again, at the
\mbox{MP2/6-31++G(d,p)} level of the theory, taking \mbox{$\sim$ 140}
days of CPU time.

Eqs.~\ref{eq:Ssk2} and ~\ref{eq:Srk2} in sec.~\ref{subsec:externals}
have been used to calculate the kinetic entropy terms associated to
the determinants of the mass-metric tensors $G$ and $g$, respectively.
The quantities in eq.~\ref{eq:Ssk2}, being simply internal
coordinates, have been directly extracted from the GAMESS output files
via automated Perl scripts. On the other hand, in order to calculate
the matrix $g_{2}$ in eq.~\ref{eq:g2} that appears in the kinetic
entropy of the classical rigid model, the Euclidean coordinates
$\vec{x}_{\alpha}^{\,\prime}$ of the 16 atoms in the reference frame
fixed in the system, as well as their derivatives with respect to
$q^{i} \equiv (\phi,\psi)$, must be computed. For this, two additional
12x12 grids as the one described above have been computed; one of them
displaced 2$^{o}$ in the positive $\phi$-direction and the other one
displaced 2$^{o}$ in the positive $\psi$-direction. This has been
done, again, at the \mbox{MP2/6-31++G(d,p)} level of the theory,
starting from the optimized structures found in the computation of the
PES described above and taking \mbox{$\sim$ 75} days of CPU time each
grid. Using the values of the positions $\vec{x}_{\alpha}^{\,\prime}$
in these two new grids and also in the original one, the derivatives
of these quantities with respect to the angles $\phi$ and $\psi$,
appearing in $g_{2}$, have been numerically obtained as finite
differences.

The three calculations have been repeated for six special points in
the Ramachandran space that correspond to important elements of
secondary structure (see sec.~\ref{sec:results}), the total CPU time
needed for computing all correcting terms at these points has been
\mbox{$\sim$ 16} days. A total of \mbox{$\sim$ 406} days of CPU time
has been needed to perform the whole study at the
\mbox{MP2/6-31++G(d,p)} level of the theory.

Finally, we have repeated all the calculations at the
\mbox{HF/6-31++G(d,p)} level of the theory in order to investigate if
this less demanding method (\mbox{$\sim$ 10} days for the PES,
\mbox{$\sim$ 8} days for the Hessians, \mbox{$\sim$ 10} days for each
displaced grid, \mbox{$\sim$ 2} days for the special secondary
structure points, being a total of \mbox{$\sim$ 40} days of CPU time)
may be used instead of MP2 in further studies.

\section{Results}
\label{sec:results}

In table~\ref{tab:sizes}, the maximum variation, the average and the
standard deviation in the 12x12 grid defined in the Ramachandran space
of the protected dipeptide HCO-{\small L}-Ala-NH$_2$ are shown for the
three energy surfaces, $V_{\Sigma}$, $F_{\mathrm{s}}$ and
$F_{\mathrm{r}}$ (see eqs.~\ref{eq:Ss} and \ref{eq:Sr}), for the
three correcting terms, $-TS_{\mathrm{s}}^{\mathrm{k}}$,
$-TS_{\mathrm{s}}^{\mathrm{c}}$, and $-TS_{\mathrm{r}}^{\mathrm{k}}$
and for the Fixman's compensating potential $V_{\mathrm{F}}$ (see
eq.~\ref{eq:VF}). All the functions have been referenced to zero
in the grid.

\begin{table}[!ht]
\begin{center}
\begin{tabular}{c@{\hspace{30pt}}r@{\hspace{20pt}}r@{\hspace{20pt}}r@{\hspace{30pt}}r@{\hspace{20pt}}r@{\hspace{20pt}}r}
\hline\\[-11pt]
\hline\\[-8pt]
 & \multicolumn{3}{c}{MP2/6-31++G(d,p)} \rule{30pt}{0pt} & 
   \multicolumn{3}{c}{HF/6-31++G(d,p)} \\[3pt]
  & Max.$^{a}$ & Ave.$^{b}$ & Std.$^{c}$ &
    Max.$^{a}$ & Ave.$^{b}$ & Std.$^{c}$ \\[3pt]
\hline\\[-11pt]
\hline\\[-6pt]
$V_{\Sigma}$                    & 21.64 & 6.76 & 3.88 & 23.62 & 6.92 & 4.35 \\
$F_{\mathrm{s}}$                & 21.43 & 6.47 & 3.93 & 23.78 & 7.17 & 4.38 \\
$F_{\mathrm{r}}$           & 21.09 & 6.46 & 3.82 & 23.09 & 6.76 & 4.31 \\[6pt]
$-TS_{\mathrm{s}}^{\mathrm{k}}$ &  0.24 & 0.09 & 0.05 &  0.23 & 0.09 & 0.04 \\
$-TS_{\mathrm{s}}^{\mathrm{c}}$ &  1.67 & 0.98 & 0.32 &  1.34 & 0.63 & 0.30 \\
$-TS_{\mathrm{r}}^{\mathrm{k}}$ &  0.81 & 0.37 & 0.12 &  0.75 & 0.38 & 0.12 \\
$V_{\mathrm{F}}$           &  1.68 & 0.89 & 0.30 &  1.35 & 0.55 & 0.27 \\[3pt]
\hline\\[-11pt]
\hline
\end{tabular}
\end{center}
\caption{\label{tab:sizes}{\small $^{a}$Maximum variation,
$^{b}$average and $^{c}$standard deviation in the 12x12 grid defined
in the Ramachandran space of the protected dipeptide HCO-{\small
L}-Ala-NH$_2$ for the three energy surfaces, $V_{\Sigma}$,
$F_{\mathrm{s}}$ and $F_{\mathrm{r}}$, the three correcting terms,
$-TS_{\mathrm{s}}^{\mathrm{k}}$, $-TS_{\mathrm{s}}^{\mathrm{c}}$, and
$-TS_{\mathrm{r}}^{\mathrm{k}}$ and the Fixman's compensating
potential $V_{\mathrm{F}}$. The results at both MP2/6-31++G(d,p) and
HF/6-31++G(d,p) levels of the theory are presented and all the
functions have been referenced to zero in the grid. The units used
are kcal/mol.}}
\end{table}

\begin{figure}[!ht]
\begin{center}
\epsfig{file=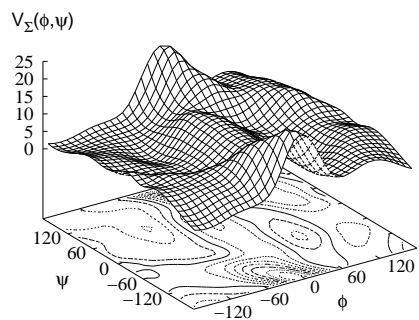,width=7cm}
\end{center}
\caption{\label{fig:pes}{\small Potential Energy Surface (PES) of the
model dipeptide HCO-{\small L}-Ala-NH$_2$, computed at the
MP2/6-31++G(d,p) level of the theory. The surface has been referenced
to zero and smoothed with bicubic splines for visual convenience. The
units in the z-axis are kcal/mol.}}
\end{figure}

In fig.~\ref{fig:pes}, the Potential Energy Surface $V_{\Sigma}$, at
the MP2/6-31++G(d,p) level of the theory, is depicted with the
reference set to zero for visual
convenience\footnote{\label{foot:referenceMP2}At the level of the
theory used in the calculations, the minimum of
$V_{\Sigma}(\phi,\psi)$ in the grid is -416.0733418995
hartree.}. Neither the surfaces defined by $F_{\mathrm{s}}$ and
$F_{\mathrm{r}}$ at the MP2/6-31++G(d,p) level of the theory nor the
three energy surfaces $V_{\Sigma}$, $F_{\mathrm{s}}$ and
$F_{\mathrm{r}}$ at HF/6-31++G(d,p) are shown graphically since they
are visually very similar to the surface in fig.~\ref{fig:pes}.

\begin{figure}[!ht]
\begin{center}
\epsfig{file=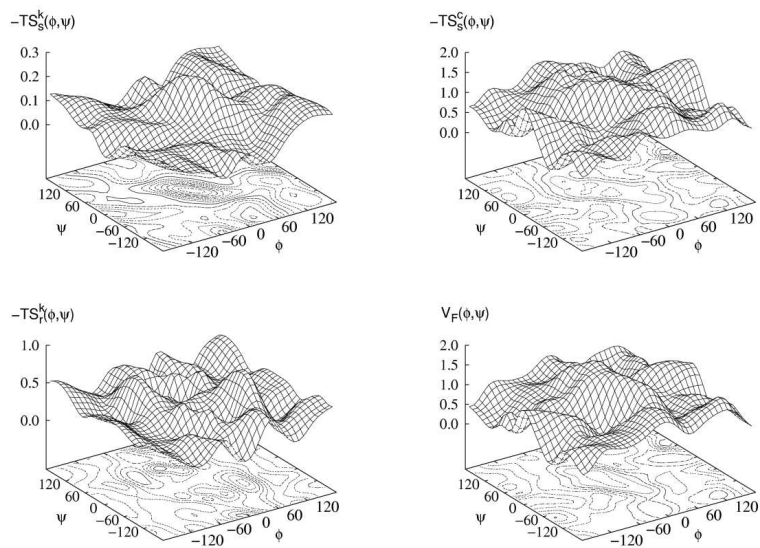,width=11cm}
\end{center}
\caption{\label{fig:correcting_terms}{\small Ramachandran plots of the
correcting terms appearing in eqs.~\ref{eq:Ss} and \ref{eq:Sr},
together with the Fixman's compensating potential defined in
eq.~\ref{eq:VF}, computed at the MP2/6-31++G(d,p) level of the
theory in the model dipeptide HCO-{\small L}-Ala-NH$_2$. The surfaces
have been referenced to zero and smoothed with bicubic splines for
visual convenience. The units in the z-axes are kcal/mol.}}
\end{figure}

In fig.~\ref{fig:correcting_terms}, the three correcting terms,
$-TS_{\mathrm{s}}^{\mathrm{k}}$, $-TS_{\mathrm{s}}^{\mathrm{c}}$ and
$-TS_{\mathrm{r}}^{\mathrm{k}}$ and the Fixman's compensating
potential $V_{\mathrm{F}}$, at the MP2/6-31++G(d,p) level of the
theory, are depicted with the reference set to zero. The analogous
surfaces at the HF/6-31++G(d,p) level of the theory are visually very
similar to the ones in fig.~\ref{fig:correcting_terms} and have been
therefore omitted.

From the results presented, one may conclude that, although the
conformational dependence of the correcting terms
$-TS_{\mathrm{s}}^{\mathrm{k}}$, $-TS_{\mathrm{s}}^{\mathrm{c}}$ and
$-TS_{\mathrm{r}}^{\mathrm{k}}$ is more than an order of magnitude
smaller than the conformational dependence of the Potential Energy
Surface $V_{\Sigma}$ in the worst case, if \emph{chemical accuracy}
(typically defined in the field of ab initio quantum chemistry as
\mbox{$1$ kcal/mol} \cite{PE:Day2003PRL}) is sought, they may be
relevant. In fact, they are of the order of magnitude of the
differences between the energy surfaces $V_{\Sigma}$, $F_{\mathrm{s}}$
and $F_{\mathrm{r}}$ calculated at MP2/6-31++G(d,p) and the ones
calculated at HF/6-31++G(d,p).

For the same reasons, we may conclude that, if ab initio derived
potentials are used to carry out Molecular Dynamics simulations of
peptides, the Fixman's compensating potential $V_{\mathrm{F}}$ should
be included. Finally, regarding the relative importance of the
different correcting terms $-TS_{\mathrm{s}}^{\mathrm{k}}$,
$-TS_{\mathrm{s}}^{\mathrm{c}}$ and $-TS_{\mathrm{r}}^{\mathrm{k}}$,
the results in table~\ref{tab:sizes} suggest that the less important
one is the kinetic entropy $-TS_{\mathrm{s}}^{\mathrm{k}}$ of the
stiff case (related to the determinant of the mass-metric tensor $G$)
and that the most important one is the one related to the determinant
of the Hessian matrix $\mathcal{H}$ of the constraining part of the
potential, i.e., the conformational entropy
$-TS_{\mathrm{s}}^{\mathrm{k}}$. The first conclusion is in agreement
with the approximations typically made in the literature, the second
one, however, is not (see the Appendix).

Now, although the relative sizes of the conformational dependence of
the different terms may be indicative of their importance, the degree
of correlation among the surfaces is also relevant (see
table~\ref{tab:correlations}). Hence, in order to arrive to more
precise conclusions, we reexamine here the results using a physically
meaningful criterium to compare potential energy functions that has
been introduced in ref.~\citen{PE:Alo2006JCC}. The \emph{distance},
denoted by $d_{12}$, between any two different potential energy
functions, $V_{1}$ and $V_{2}$, is an statistical quantity that, from
a working set of conformations (in this case, the 144 points of the
grid), measures the typical error that one makes in the \emph{energy
differences} if $V_{2}$ is used instead of $V_{1}$, admitting a linear
rescaling.

\begin{table}[!ht]
\begin{center}
\begin{tabular}{c@{\hspace{20pt}}c@{\hspace{20pt}}c@{\hspace{20pt}}rrrr}
\hline\\[-11pt]
\hline\\[-8pt]
Corr.$^{a}$ & ${V_{1}}^{b}$ & ${V_{2}}^{c}$ &
 \multicolumn{1}{c}{${d_{12}}^{d}$} &
 \multicolumn{1}{c}{${N_{\mathrm{res}}}^{e}$} &
 \multicolumn{1}{c}{${b_{12}}^{f}$} &
 \multicolumn{1}{c}{${r_{12}}^{g}$} \\[3pt]
\hline\\[-11pt]
\hline\\[-6pt]
\multicolumn{7}{c}{MP2/6-31++G(d,p)} \\[4pt]
$- TS_{\mathrm{s}}^{\mathrm{k}} - TS_{\mathrm{s}}^{\mathrm{c}}$ &
 $F_{\mathrm{s}}$ & $V_{\Sigma}$ &
 0.74 $RT$ & 1.82 & 0.98 & 0.9967 \\
$- TS_{\mathrm{s}}^{\mathrm{c}}$ &
 $F_{\mathrm{s}}$ & $V_{\Sigma} - TS_{\mathrm{s}}^{\mathrm{k}}$ &
 0.74 $RT$ & 1.83 & 0.98 & 0.9967 \\
$- TS_{\mathrm{s}}^{\mathrm{k}}$ &
 $F_{\mathrm{s}}$ & $V_{\Sigma} - TS_{\mathrm{s}}^{\mathrm{c}}$ &
 0.11 $RT$ & 80.45 & 1.00 & 0.9999 \\[6pt]
$- TS_{\mathrm{r}}^{\mathrm{k}}$ &
 $F_{\mathrm{r}}$ & $V_{\Sigma}$ &
 0.29 $RT$ & 11.62 & 1.01 & 0.9995 \\[6pt]
$V_{\mathrm{F}}$ &
 $F_{\mathrm{s}}$ & $F_{\mathrm{r}}$ &
 0.67 $RT$ & 2.24 & 0.97 & 0.9972 \\[3pt]
\hline\\[-6pt]
\multicolumn{7}{c}{HF/6-31++G(d,p)} \\[4pt]
$- TS_{\mathrm{s}}^{\mathrm{k}} - TS_{\mathrm{s}}^{\mathrm{c}}$ &
 $F_{\mathrm{s}}$ & $V_{\Sigma}$ &
 0.73 $RT$ & 1.90 & 0.99 & 0.9975 \\
$- TS_{\mathrm{s}}^{\mathrm{c}}$ &
 $F_{\mathrm{s}}$ & $V_{\Sigma} - TS_{\mathrm{s}}^{\mathrm{k}}$ &
 0.71 $RT$ & 2.00 & 0.99 & 0.9976 \\
$- TS_{\mathrm{s}}^{\mathrm{k}}$ &
 $F_{\mathrm{s}}$ & $V_{\Sigma} - TS_{\mathrm{s}}^{\mathrm{c}}$ &
 0.10 $RT$ & 90.99 & 1.00 & 0.9999 \\[6pt]
$- TS_{\mathrm{r}}^{\mathrm{k}}$ &
 $F_{\mathrm{r}}$ & $V_{\Sigma}$ &
 0.26 $RT$ & 14.83 & 1.01 & 0.9997 \\[6pt]
$V_{\mathrm{F}}$ &
 $F_{\mathrm{s}}$ & $F_{\mathrm{r}}$ &
 0.61 $RT$ & 2.69 & 0.98 & 0.9982 \\[3pt]
\hline\\[-6pt]
\multicolumn{7}{c}{MP2/6-31++G(d,p) vs. HF/6-31++G(d,p)} \\[4pt]
& $V_{\Sigma}$ & $V_{\Sigma}$ &
 1.25 $RT$ & 0.64 & 1.12 & 0.9925 \\
& $F_{\mathrm{s}}$ & $F_{\mathrm{s}}$ &
 1.18 $RT$ & 0.72 & 1.11 & 0.9934 \\
& $F_{\mathrm{r}}$ & $F_{\mathrm{r}}$ &
 1.18 $RT$ & 0.72 & 1.12 & 0.9932 \\[3pt]
\hline\\[-11pt]
\hline
\end{tabular}
\end{center}
\caption{\label{tab:distances}{\small Comparison of different energy
surfaces involved in the study of the constrained equilibrium of the
protected dipeptide HCO-{\small L}-Ala-NH$_2$. $^{a}$Correcting term
whose importance is measured in the corresponding row, $^{b}$reference
potential energy $V_{1}$ (the ``correct'' one, the one containing the
correcting term), $^{c}$approximated potential energy $V_{2}$ (i.e,
$V_{1}$ minus the correcting term in column $a$), $^{d}$statistical
distance between $V_{1}$ and $V_{2}$ (see Alonso J. L. and Echenique
P., \emph{J. Comp. Chem.}  {\bf 27} (2006) 238--252), $^{e}$maximum
number of residues in a polypeptide potential up to which the
correcting term in column $a$ may be omitted, $^{f}$slope of the
linear rescaling between $V_{1}$ and $V_{2}$ and $^{g}$Pearson's
correlation coefficient. All quantities are dimensionless, except for
$d_{12}$ which is given in units of the thermal energy $RT$ at
\mbox{$300^{o}$ K}.}}
\end{table}

In table~\ref{tab:distances}, which contains the central results of
this work, the distances between some of the energy surfaces that play
a role in the problem are shown. We present the result in units of
$RT$ (at \mbox{$300^{o}$ K}, where $RT\simeq 0.6$ kcal/mol) because it
has been argued in ref.~\citen{PE:Alo2006JCC} that, if the distance
between two different approximations of the energy of the same system
is less than $RT$, one may safely substitute one by the other without
altering the relevant physical properties. Moreover, if one assumes
that the effective energies compared will be used to construct a
polypeptide potential and that it will be designed as simply the sum
of mono-residue ones (making each term suitably depend on different
pairs of Ramachandran angles), then, the number $N_{\mathrm{res}}$ of
residues up to which one may go keeping the distance between the two
approximations of the the $N$-residue potential below $RT$ is (see
eq.~23 in ref.~\citen{PE:Alo2006JCC}):

\begin{equation}
\label{eq:Nres}
N_{\mathrm{res}}=\left ( \frac{RT}{d_{12}} \right )^{2} \ .
\end{equation}

This number is also shown in table~\ref{tab:distances}, together
with the slope $b_{12}$ of the linear rescaling between $V_{1}$ and
$V_{2}$ and the Pearson's correlation coefficient \cite{PE:Dob1991BOOK},
denoted by $r_{12}$.

\begin{table}[!ht]
\begin{center}
\begin{tabular}{ccc@{\hspace{30pt}}r}
\hline\\[-11pt]
\hline\\[-8pt]
${V_{1}}^{a}$ & & ${V_{2}}^{b}$ & \multicolumn{1}{c}{${r_{12}}^{c}$} \\[3pt]
\hline\\[-11pt]
\hline\\[-6pt]
\multicolumn{4}{c}{MP2/6-31++G(d,p)} \\[4pt]
$V_{\Sigma}$ & vs. & $-TS_{\mathrm{s}}^{\mathrm{c}}$ &  0.1572 \\
$V_{\Sigma}$ & vs. & $-TS_{\mathrm{s}}^{\mathrm{k}}$ & -0.0008 \\
$V_{\Sigma}$ & vs. & $-TS_{\mathrm{r}}^{\mathrm{k}}$ & -0.3831 \\
$V_{\Sigma}$ & vs. & $V_{\mathrm{F}}$                &  0.3334 \\[3pt]
\hline\\[-6pt]
\multicolumn{4}{c}{HF/6-31++G(d,p)} \\[4pt]
$V_{\Sigma}$ & vs. & $-TS_{\mathrm{s}}^{\mathrm{c}}$ &  0.0682 \\
$V_{\Sigma}$ & vs. & $-TS_{\mathrm{s}}^{\mathrm{k}}$ &  0.0897 \\
$V_{\Sigma}$ & vs. & $-TS_{\mathrm{r}}^{\mathrm{k}}$ & -0.3544 \\
$V_{\Sigma}$ & vs. & $V_{\mathrm{F}}$                &  0.2404 \\[3pt]
\hline\\[-6pt]
\multicolumn{4}{c}{MP2/6-31++G(d,p) vs. HF/6-31++G(d,p)} \\[4pt]
$-TS_{\mathrm{s}}^{\mathrm{c}}$ & vs. & 
$-TS_{\mathrm{s}}^{\mathrm{c}}$ &  0.9136 \\
$-TS_{\mathrm{s}}^{\mathrm{k}}$ & vs. & 
$-TS_{\mathrm{s}}^{\mathrm{k}}$ &  0.9808 \\
$-TS_{\mathrm{r}}^{\mathrm{k}}$ & vs. & 
$-TS_{\mathrm{r}}^{\mathrm{k}}$ &  0.9316 \\
$V_{\mathrm{F}}$ & vs. & 
$V_{\mathrm{F}}$                &  0.9217 \\[3pt]
\hline\\[-11pt]
\hline
\end{tabular}
\end{center}
\caption{\label{tab:correlations}{\small Correlation between the
different correcting terms involved in the study of the constrained
equilibrium of the protected dipeptide HCO-{\small
L}-Ala-NH$_2$. $^{a}$Reference potential energy, $^{b}$approximated
potential energy, $^{c}$Pearson's correlation coefficient.}}
\end{table}

The results at both MP2/6-31++G(d,p) and HF/6-31++G(d,p) levels of the
theory are presented. The first three rows in each of the first two
blocks are related to the classical stiff model, the next row to the
classical rigid model and the last one in each block to the comparison
between the two models. The third block in the table is associated to
the comparison between the two different levels of the theory used.

The $F_{\mathrm{s}}$ vs. $V_{\Sigma}$ row (in the first two blocks)
assess the importance of the two correcting terms,
$-TS_{\mathrm{s}}^{\mathrm{k}}$ and $-TS_{\mathrm{s}}^{\mathrm{c}}$,
in the stiff case. The result $d_{12}=0.74 RT$ indicates that, for the
alanine dipeptide, $V_{\Sigma}$ may be used as an approximation of
$F_{\mathrm{s}}$ with caution if accurate results are sought. In fact,
the low value of $N_{\mathrm{res}}=1.82 < 2$ shows that, \emph{if we
wanted to describe a 2-residue peptide omitting the stiff correcting
terms, we would typically make an error greater than the thermal noise
in the energy differences.} The next two rows investigate the effect
of each one of the individual correcting terms. The conclusion that
can be extracted from them (as the relative sizes in
table~\ref{tab:sizes} already suggested) is that the conformational
entropy associated to the determinant of the Hessian matrix
$\mathcal{H}$ is much more relevant than the correcting term
$-TS_{\mathrm{s}}^{\mathrm{k}}$, related to the mass-metric tensor
$G$, \emph{allowing to drop the latter up to $\sim 80$ residues}
(according to MP2/6-31++G(d,p) calculations).  As has been already
remarked, this second conclusion is in agreement with the
approximations frequently done in the literature; however, it turns
out that the importance of the Hessian-related term has been
persistently underestimated (see the Appendix for a discussion).

The $F_{\mathrm{r}}$ vs. $V_{\Sigma}$ row, in turn, shows the data
associated to the kinetic entropy term
$-TS_{\mathrm{r}}^{\mathrm{k}}$, which is related to the determinant
of the reduced mass-metric tensor $g$ in the classical rigid
model. From the results there ($d_{12}=0.29 RT$ and
$N_{\mathrm{res}}=11.62$ at the MP2/6-31++G(d,p) level), we can
conclude that the only correction term in the rigid case is less
important than the ones in the stiff case and that \emph{$V_{\Sigma}$
may be used as an approximation of $F_{\mathrm{r}}$ for oligopeptides
of up to $\sim 12$ residues.}

The last row in each of the first two blocks in
table~\ref{tab:distances} is related to the interesting question in
Molecular Dynamics of whether or not one should include the Fixman's
compensating potential $V_{F}$ (see eq.~\ref{eq:VF}) in rigid
simulations in order to obtain the stiff equilibrium distribution,
$\exp (-\beta F_{\mathrm{s}})$, instead of the rigid one, $\exp
(-\beta F_{\mathrm{r}})$. This question is equivalent to asking
whether or not $F_{\mathrm{r}}$ is a good approximation of
$F_{\mathrm{s}}$. From the results in the table, we can conclude that
\emph{the Fixman's potential is relevant for peptides of more than 2
residues and its omission may cause an error greater than the thermal
noise in the energy differences.}

The appreciable sizes of the different correcting terms, shown
in table~\ref{tab:sizes}, together with their low correlation
with the Potential Energy Surface $V_{\Sigma}$, presented in the
first two blocks of table~\ref{tab:correlations}, explain their
considerable relevance discussed in the preceding paragraphs.

Moreover, from the comparison of the MP2/6-31++G(d,p) and the
HF/6-31++G(d,p) blocks, one can tell that \emph{the study herein
performed may well have been done at the lower level of the theory}
(if we had known) with a tenth of the computational effort (see
sec.~\ref{sec:methods}).  This fact, explained by the high
correlation, presented in the third block of
table~\ref{tab:correlations}, between the correcting terms calculated
at the two levels, is \emph{very relevant for further studies} on more
complicated dipeptides or longer chains and it indicates that the
differences in size between the different correcting terms at
MP2/6-31++G(d,p) and HF/6-31++G(d,p), which are presented in
table~\ref{tab:sizes}, are mostly due to a harmless linear scaling
effect similar to the well-known empirical scale factor frequently
used in ab initio vibrational analysis
\cite{PE:Lev1999BOOK,PE:Hal2001TCA,PE:Sco1996JPC}. This view is
supported by the data in the third block of table~\ref{tab:distances},
related to the comparison between the energy surfaces calculated at
MP2/6-31++G(d,p) and HF/6-31++G(d,p), where the slopes $b_{12}$ are
consistently larger than unity.

A last conclusion that may be extracted from the block labeled
``MP2/6-31++G(d,p) vs. HF/6-31++G(d,p)'' in table~\ref{tab:distances}
is that the typical error in the energy differences (given by the
distances $d_{12}$) produced when one reduces the level of the theory
from MP2/6-31++G(d,p) to HF/6-31++G(d,p) \emph{is comparable} (less
than twice) to the error made if the most important correcting terms
of the classical constrained models studied in this work are
dropped. This is a useful hint for researchers interested in the
conformational analysis of peptides with quantum chemistry methods
\cite{PE:Lan2005PSFB,PE:Per2003JCC,PE:Var2002JPCA,PE:Els2001CP,PE:Yu2001JMS,PE:Csa1999PBMB,PE:Bea1997JACS}
and also to those whose aim is the design and parametrization of
classical force fields from ab initio quantum mechanical calculations
\cite{PE:Mac2004JCC,PE:Bor2003JPCB,PE:Bea1997JACS}.

\begin{table}[!ht]
\begin{center}
\begin{tabular}{l@{\hspace{20pt}}rr}
\hline\\[-11pt]
\hline\\[-8pt]
 & \multicolumn{1}{c}{$\phi$} & \multicolumn{1}{c}{$\psi$} \\[3pt]
\hline\\[-11pt]
\hline\\[-6pt]
$\alpha$-helix             &  -57 & -47 \\
3$_{10}$-helix             &  -49 & -26 \\
$\pi$-helix                &  -57 & -70 \\
polyproline II             &  -79 & 149 \\
parallel $\beta$-sheet     & -119 & 113 \\
antiparallel $\beta$-sheet & -139 & 135 \\[3pt]
\hline\\[-11pt]
\hline
\end{tabular}
\end{center}
\caption{\label{tab:secondary_def}{\small Ramachandran angles (in
degrees) of some important secondary structure elements in
polypeptides. Data taken from Lesk A. M., \emph{Introduction to
Protein Architecture}, Oxford University Press, Oxford, 2001.}}
\end{table}

Finally, in order to enrich and qualify the analysis, a new
\emph{working set} of conformations, different from the 144 points of
the grid in the Ramachandran space, have been selected and the whole
study has been repeated on them. These new conformations are six
important secondary structure elements which form repetitive patterns
stabilized by hydrogen bonds in polypeptides. Their conventional names
and the corresponding values of the $\phi$ and $\psi$ angles have been
taken from ref.~\citen{PE:Les2001BOOK} and are shown in
table~\ref{tab:secondary_def}.

\begin{figure}[!ht]
\begin{center}
\epsfig{file=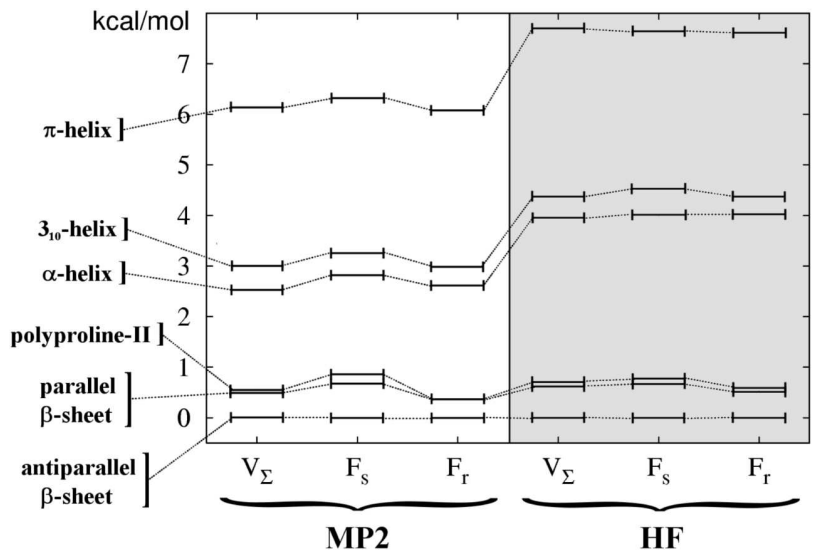,width=11cm}
\end{center}
\caption{\label{fig:secondary}{\small Relative energies of some
important elements of secondary structure for the three potentials
$V_{\Sigma}$, $F_{\mathrm{s}}$ and $F_{\mathrm{r}}$, in the model
dipeptide HCO-{\small L}-Ala-NH$_2$ and at both MP2/6-31++G(d,p) and
HF/6-31++G(d,p) levels of the theory. The energy of the antiparallel
$\beta$-sheet has been taken as reference. The units are kcal/mol.}}
\end{figure}

In fig.~\ref{fig:secondary}, the relative energies of these
conformations are shown for the three relevant potentials,
$V_{\Sigma}$, $F_{\mathrm{s}}$ and $F_{\mathrm{r}}$, at both
MP2/6-31++G(d,p) and HF/6-31++G(d,p) levels of the theory. Since the
antiparallel $\beta$-sheet is the structure with the minimum energy in
all the cases, it has been set as the reference and the rest of
energies in the figure should be regarded as relative to it.

The meaningful assessment, using the statistical distance described
above, of the typical error made in the energy differences has been
also performed on this new working set of conformations. The results
are presented in table~\ref{tab:secondary_dist}.

\begin{table}[!ht]
\begin{center}
\begin{tabular}{c@{\hspace{20pt}}c@{\hspace{20pt}}c@{\hspace{20pt}}rrrr}
\hline\\[-11pt]
\hline\\[-8pt]
Corr.$^{a}$ & ${V_{1}}^{b}$ & ${V_{2}}^{c}$ &
 \multicolumn{1}{c}{${d_{12}}^{d}$} &
 \multicolumn{1}{c}{${N_{\mathrm{res}}}^{e}$} &
 \multicolumn{1}{c}{${b_{12}}^{f}$} &
 \multicolumn{1}{c}{${r_{12}}^{g}$} \\[3pt]
\hline\\[-11pt]
\hline\\[-6pt]
\multicolumn{7}{c}{MP2/6-31++G(d,p)} \\[4pt]
$- TS_{\mathrm{s}}^{\mathrm{k}} - TS_{\mathrm{s}}^{\mathrm{c}}$ &
 $F_{\mathrm{s}}$ & $V_{\Sigma}$ &
 0.22 $RT$ & 19.72 & 0.99 & 0.9990 \\
$- TS_{\mathrm{s}}^{\mathrm{c}}$ &
 $F_{\mathrm{s}}$ & $V_{\Sigma} - TS_{\mathrm{s}}^{\mathrm{k}}$ &
 0.26 $RT$ & 14.07 & 0.98 & 0.9985 \\
$- TS_{\mathrm{s}}^{\mathrm{k}}$ &
 $F_{\mathrm{s}}$ & $V_{\Sigma} - TS_{\mathrm{s}}^{\mathrm{c}}$ &
 0.06 $RT$ & 298.13 & 1.01 & 0.9999 \\[6pt]
$- TS_{\mathrm{r}}^{\mathrm{k}}$ &
 $F_{\mathrm{r}}$ & $V_{\Sigma}$ &
 0.20 $RT$ & 25.64 & 0.99 & 0.9992 \\[6pt]
$V_{\mathrm{F}}$ &
 $F_{\mathrm{s}}$ & $F_{\mathrm{r}}$ &
 0.34 $RT$ & 8.73 & 0.99 & 0.9977 \\[3pt]
\hline\\[-6pt]
\multicolumn{7}{c}{HF/6-31++G(d,p)} \\[4pt]
$- TS_{\mathrm{s}}^{\mathrm{k}} - TS_{\mathrm{s}}^{\mathrm{c}}$ &
 $F_{\mathrm{s}}$ & $V_{\Sigma}$ &
 0.14 $RT$ & 47.94 & 1.00 & 0.9997 \\
$- TS_{\mathrm{s}}^{\mathrm{c}}$ &
 $F_{\mathrm{s}}$ & $V_{\Sigma} - TS_{\mathrm{s}}^{\mathrm{k}}$ &
 0.15 $RT$ & 46.12 & 1.00 & 0.9997 \\
$- TS_{\mathrm{s}}^{\mathrm{k}}$ &
 $F_{\mathrm{s}}$ & $V_{\Sigma} - TS_{\mathrm{s}}^{\mathrm{c}}$ &
 0.05 $RT$ & 380.30 & 1.00 & 0.9999 \\[6pt]
$- TS_{\mathrm{r}}^{\mathrm{k}}$ &
 $F_{\mathrm{r}}$ & $V_{\Sigma}$ &
 0.15 $RT$ & 41.85 & 0.99 & 0.9997 \\[6pt]
$V_{\mathrm{F}}$ &
 $F_{\mathrm{s}}$ & $F_{\mathrm{r}}$ &
 0.18 $RT$ & 30.12 & 1.01 & 0.9996 \\[3pt]
\hline\\[-6pt]
\multicolumn{7}{c}{MP2/6-31++G(d,p) vs. HF/6-31++G(d,p)} \\[4pt]
& $V_{\Sigma}$ & $V_{\Sigma}$ &
 0.77 $RT$ & 1.68 & 1.28 & 0.9929 \\
& $F_{\mathrm{s}}$ & $F_{\mathrm{s}}$ &
 0.77 $RT$ & 1.69 & 1.26 & 0.9928 \\
& $F_{\mathrm{r}}$ & $F_{\mathrm{r}}$ &
 0.71 $RT$ & 1.96 & 1.28 & 0.9939 \\[3pt]
\hline\\[-11pt]
\hline
\end{tabular}
\end{center}
\caption{\label{tab:secondary_dist}{\small Comparison of different
approximations to the energies of some important elements of secondary
structure (see table~\ref{tab:secondary_def}) in the study of the
constrained equilibrium of the protected dipeptide HCO-{\small
L}-Ala-NH$_2$. See the caption of table~\ref{tab:distances} for an
explanation of the keys in the different columns.}}
\end{table}

The distances between the free energies, $F_{\mathrm{s}}$ and
$F_{\mathrm{r}}$, and their corresponding approximations obtained
dropping the correcting entropies, $-TS_{\mathrm{s}}^{\mathrm{k}}$,
$-TS_{\mathrm{s}}^{\mathrm{c}}$ and $-TS_{\mathrm{r}}^{\mathrm{k}}$,
or the Fixman's compensating potential $V_{\mathrm{F}}$, in the
first two blocks of the table, are \emph{consistently smaller than the
ones found in the study of the grid defined in the whole
Ramachandran space} (cf. table~\ref{tab:distances}). And so
are the distances between the three relevant potentials,
$V_{\Sigma}$, $F_{\mathrm{s}}$ and $F_{\mathrm{r}}$, calculated
at the MP2/6-31++G(d,p) and HF/6-31++G(d,p) levels of the theory.

Although the distance $d_{12}$ used is a statistical quantity and,
therefore, one must be cautious when working with such a small set of
conformations (of size six, in this case), the conclusion drawn from
this second part of the study is that, if one is interested only in
the ``lower region'' of the Ramachandran surface, where the typical
secondary structure elements lie, then, \emph{one may safely neglect
the conformational dependence of the different correcting terms
appearing in the study of the constrained equilibrium of peptides}. At
least, up to oligopeptides (poly-alanines) of $\sim$ 10 residues in
the worst case (the neglect of the Fixman's compensating potential
$V_{\mathrm{F}}$ in the $F_{\mathrm{s}}$ vs. $F_{\mathrm{r}}$
comparison at MP2/6-31++G(d,p)).

This difference between the two working set of conformations may be
explained looking at one of the ways of expressing the statistical
distance used (see eq.~12a in ref.~\citen{PE:Alo2006JCC}):

\begin{equation}
\label{eq:d}
d_{12} = \sqrt{2}\,{\sigma}_{2}(1-r_{12}^{2})^{1/2} \  ,
\end{equation}

where $r_{12}$ is the Pearson's correlation coefficient between
the potential energies denoted by $V_{1}$ and $V_{2}$ and
${\sigma}_{2}$ is the standard deviation in the values of $V_{2}$
on the relevant working set of conformations.

This last quantity, ${\sigma}_{2}$, is the responsible of the
differences between tables~\ref{tab:distances} and
\ref{tab:secondary_dist}, since the set of conformations comprised by
the six secondary structure elements in table~\ref{tab:secondary_def}
spans a smaller energy range than the whole Potential Energy Surface
in fig.~\ref{fig:pes} (or $F_{\mathrm{s}}$, or $F_{\mathrm{r}}$, which
have very similar variations). Accordingly, the dispersion in the
energy values is smaller: $\sigma_{2} \simeq$ 2 kcal/mol in the case
of the secondary structure elements and $\sigma_{2} \simeq$ 4 kcal/mol
for the grid in the whole Ramachandran space (see
table~\ref{tab:sizes}).  Since the correlation coefficient in both
cases are of similar magnitude, the differences in $\sigma_{2}$
produce a smaller distance $d_{12}$ for the second set of
conformations studied, i.e., a smaller typical error made in the
energy differences when omitting the correcting terms derived from the
consideration of constraints.

To end this section, we remark that, although this ``lower region'' of
the Ramachandran space contains the most relevant secondary structure
elements (which are also the most commonly found in experimentally
resolved native structures of proteins
\cite{PE:Cha2001PBMB,PE:Ber2000NAR,PE:Gun1996JMB,PE:Cre1992BOOK}) and
may be the only region explored in the dynamical or thermodynamical
study of small peptides, if the aim is the design of effective
potentials for computer simulation of polypeptides
\cite{PE:Mac2004JCC,PE:Bor2003JPCB,PE:Bea1997JACS}, then, some caution
is recommended, since long-range interactions in the sequence may
temporarily compensate local energy penalizations and the higher
regions of the energy surfaces studied could be important in
transition states or in some relevant dynamical paths of the system.

In the following section, the many results discussed in the preceding
paragraphs are summarized.

\section{Conclusions}
\label{sec:conclusions}

In this work, the theory of classical constrained equilibrium has been
collected for the stiff and rigid models. The pertinent correcting
terms, which may be regarded as effective entropies, as well as the
Fixman's compensating potential, have been derived and theoretically
discussed (see eqs.~\ref{eq:Ss}, \ref{eq:Sr} and \ref{eq:VF}, together
with the formulae in sec.~\ref{subsec:externals}). In addition, the
common approximation of considering that, for typical internals, the
equilibrium values of the hard coordinates do not depend on the soft
ones, has also been discussed and related to the rest of
simplifications. The treatment of the assumptions in the
literature is thoroughly reviewed and discussed in the Appendix.

In the central part of the work (sec.~\ref{sec:results}), the
relevance of the different correcting terms has been assessed in the
case of the model dipeptide HCO-{\small L}-Ala-NH$_2$, with quantum
mechanical calculations including electron correlation. Also, the
possibility of performing analogous studies at the less demanding
Hartree-Fock level of the the theory has been investigated. The
results found are summarized in the following points:

\begin{itemize}
\item \emph{In Monte Carlo simulations of the classical stiff model}
 at room temperature, the effective entropy
 $-TS_{\mathrm{s}}^{\mathrm{k}}$, associated to the determinant of the
 mass-metric tensor $G$, may be neglected for peptides of up to
 \mbox{$\sim$ 80} residues. Its maximum variation in the Ramachandran
 space is 0.24 kcal/mol.
\item \emph{In Monte Carlo simulations of the classical stiff model}
 at room temperature, the effective entropy
 $-TS_{\mathrm{s}}^{\mathrm{c}}$, associated to the determinant of the
 Hessian $\mathcal{H}$ of the constraining part of the potential,
 should be included for peptides of more than 2 residues. Its maximum
 variation in the Ramachandran space is \mbox{1.67 kcal/mol}.
\item \emph{In Monte Carlo simulations of the classical rigid model}
 at room temperature, the effective entropy
 $-TS_{\mathrm{r}}^{\mathrm{k}}$, associated to the determinant of the
 reduced mass-metric tensor $g$, may be neglected for peptides of up
 to \mbox{$\sim$ 12} residues. Its maximum variation in the
 Ramachandran space is \mbox{0.81 kcal/mol}.
\item \emph{In rigid Molecular Dynamics simulations intended to yield
 the stiff equilibrium distribution} at room temperature, the Fixman's
 compensating potential $V_{\mathrm{F}}$ should be included for
 peptides of more than 2 residues. Its maximum variation in the
 Ramachandran space is \mbox{1.68 kcal/mol}.
\item If the assumption that only the more stable region of the
 Ramachandran space, where the principal elements of secondary
 structure lie, is relevant, then, \emph{the importance of the
 correcting terms decreases and the limiting number of residues in a
 polypeptide potential up to which they may be omitted is
 approximately four times larger in each of the previous points}.
\item In both cases (i.e., either if the whole Ramachandran space is
 considered relevant, or only the lower region), \emph{the errors made
 if the most important correcting terms are neglected are of the same
 order of magnitude as the errors due to a decrease in the level of
 theory from} MP2/6-31++G(d,p) \emph{to} HF/6-31++G(d,p).
\item \emph{The whole study of the relevance of the different
 correcting terms (or future analogous investigations) may be
 performed at the} HF/6-31++G(d,p) \emph{level of the theory},
 yielding very similar results to the ones obtained at
 MP2/6-31++G(d,p) and using a tenth of the computational effort.
\end{itemize}

To end this discussion, some qualifications should be made. On one
hand, the conclusions above refer to the case in which a classical
potential \emph{directly extracted} from the quantum mechanical
(Born-Oppenheimer) one is used; for the considerably simpler force
fields typically used for macromolecular simulations, the study should
be repeated and different results may be obtained. On the other hand,
the investigation performed in this work has been done in one of the
simplest dipeptides; both its isolated character and the relatively
small size of its side chain play a role in the results
obtained. Hence, for bulkier residues included in polypeptides, these
conclusions should be approached with caution and much interesting
work remains to be done.

\section*{Acknowledgments}

\hspace{0.5cm} We would like to thank F. Falceto and V. Laliena for
illuminating discussions and also the reviewers of the manuscript for
much useful suggestions. The numerical calculations have been
performed at the BIFI computing facilities. We thank I. Campos, for
the invaluable CPU time and the efficiency at solving the problems
encountered.

This work has been supported by the Arag\'on Government
(``Biocomputaci\'on y F\'{\i}sica de Sistemas Complejos'' group) and
by the research grants MEC (Spain) \mbox{FIS2004-05073} and
\mbox{FPA2003-02948}, and MCYT (Spain) \mbox{BFM2003-08532}. P.
Echenique and I. Calvo are supported by MEC (Spain) FPU grants.

\section*{Appendix}
\label{sec:appendix}

Many approximations may be done to simplify the calculation of
the different correcting terms introduced in the previous
subsections. The most frequently found in the literature are
the following three:

\begin{enumerate}
\item[(i)] To neglect the conformational dependence of $\det G$.
\item[(ii)] To neglect the conformational dependence of $\det \mathcal{H}$.
\item[(iii)] To assume that the hard coordinates are constant, i.e, that
 the $f^{I}(q^{i})$ in eq.~\ref{eq:real_constraints} do not depend
 on the soft coordinates $q^{i}$.
\end{enumerate}

The conformational dependence of $\det g$ is customarily regarded as
important since it was shown to be non-negligible even for simple
systems some decades ago
\cite{PE:Per1985MM,PE:Pea1979JCP,PE:Ral1979JFM,PE:Cha1979JCP}
(normally in an indirect way, while studying the influence of the
Fixman's compensating potential in eq.~\ref{eq:VF}; see discussion
below). With this same aim, Patriciu et al. \cite{PE:Pat2004JCP} have
very recently measured the conformational dependence of $\det g$ for a
serial polymer with fixed bond lengths and bond angles (in the
approximation (iii)), showing that it is non-negligible and suggesting
that it may be so also for more general systems.

Note that, if approximations (i) and (ii) are assumed, then the
Fixman's potential depends only on $\det g$. In fact, whereas in the
general case the Fixman's compensating potential cannot be simplified
beyond the expression in eq.~\ref{eq:VF}, if one assumes
approximation (iii), then the reduced mass-metric tensor $g$ turns out
to be the subblock of $G$ with soft indices and, in this case, the
quotient $\det G / \det g$ has been shown to be equal to $1 / \det h$
by Fixman \cite{PE:Fix1974PNAS}, where $h$ denotes the subblock of
$G^{-1}$ with hard indices, i.e.,

\begin{equation}
\label{eq:h}
h^{IJ}(q^{\mu}):=\sum_{\sigma=1}^{N}\frac{\partial q^{I}}
                                         {\partial x^{\sigma}}
                                    \frac{1}{m_{\sigma}}
                                    \frac{\partial q^{J}}
                                         {\partial x^{\sigma}} \ .
\end{equation}

This result has been extensively used in the literature
\cite{PE:Pas2002JCP,PE:Alm1990MP,PE:Per1985MM,PE:Pea1979JCP,PE:Cha1979JCP,PE:Fix1978JCP},
since each of the internal coordinates $q^{a}$ typically used in
macromolecular simulations only involves a small number of atoms, thus
rendering the matrix $h$ above sparse and allowing for efficient
algorithms to be used in order to find its determinant.

Now, although $\det g$ is customarily regarded as important, the
conformational variations of $\det G$ are almost unanimously neglected
(approximation (i)) in the literature \cite{PE:Per1994JCC,PE:Go1976MM}
and may only be said to be indirectly included in $h$ by the authors
that use the expression above
\cite{PE:Pat2004JCP,PE:Pas2002JCP,PE:Alm1990MP,PE:Per1985MM,PE:Pea1979JCP,PE:Cha1979JCP}.
This is mainly due to the fact, reported by G\=o and Scheraga
\cite{PE:Go1976MM} and, before, by Volkenstein
\cite{PE:Vol1959BOOKpp}, that $\det G$ in a serial polymer may be
expressed as in eq.~\ref{eq:detG_sasmic}, being independent of the
dihedral angles (which are customarily taken as the soft
coordinates). If one also assumes approximation (iii), which, as will
be discussed later, is very common, then $\det G$ is a constant for
every conformation of the molecule.

Probably due to computational considerations, but also sometimes to
the use of a formulation of the stiff case based on delta functions
\cite{PE:Sch2003MP}, the conformational dependence of $\det
\mathcal{H}$ is almost unanimously neglected (approximation (ii)) in
the literature
\cite{PE:All2005BOOK,PE:Pat2004JCP,PE:Fre2002BOOK,PE:Den2000MP,PE:Per1994JCC,PE:Ber1984BOOK,PE:Go1976MM,PE:Fix1974PNAS}.
Only a few authors include this term in different stages of the
reasoning
\cite{PE:Mor2004ACP,PE:Den2000MP,PE:Den1998JCP,PE:Ber1983BOOK,PE:Ral1979JFM,PE:Hel1979JCP,PE:Go1976MM},
most of them only to argue later that it is negligible.

Although for some simple ad hoc designed potentials that lack
long-range terms \cite{PE:Alm1990MP,PE:Per1985MM,PE:Pea1979JCP}, the
aforementioned simplifying assumptions and the ones that will be
discussed in the following paragraphs may be exactly fulfilled, in the
case of the potential energies used in force fields for macromolecular
simulation
\cite{PE:Mac1998BOOK,PE:Bro1983JCC,PE:VGu1982MM,PE:Cor1995JACS,PE:Pea1995CPC,PE:Jor1988JACS,PE:Jor1996JACS,PE:Hal1996JCCa,PE:Hal1996JCCb,PE:Hal1996JCCc,PE:Hal1996JCCd,PE:Hal1996JCCe},
they are not. The typical energy function in this case, has the form

\begin{eqnarray}
\label{eq:Vff}
V_{\mathrm{ff}}(q^{a}) & := &
   \frac{1}{2}\sum_{\alpha = 1}^{N_{r}}
      K_{r_{\alpha}}(r_{\alpha}-r_{\alpha}^{0})^{2} + 
   \frac{1}{2}\sum_{\alpha = 1}^{N_{\theta}}
      K_{\theta_{\alpha}}(\theta_{\alpha}-\theta_{\alpha}^{\,0})^{2} + 
 \nonumber \\
& + & V_{\mathrm{ff}}^{\mathrm{tors}}(\phi_{\alpha}) +
      V_{\mathrm{ff}}^{\mathrm{long-range}}(q^{a}) \ ,
\end{eqnarray}

where $r_{\alpha}$ are bond lengths, $\theta_{\alpha}$ are bond
angles, $\phi_{\alpha}$ are dihedral angles and, for the sake of
simplicity, no harmonic terms have been assumed for out-of-plane
angles or for hard dihedrals (such as the peptide bond $\omega$).
$N_{r}$ is the number of bond lengths, $N_{\theta}$ the number of bond
angles and the quantities $K_{r_{\alpha}}$, $K_{\theta_{\alpha}}$,
$r_{\alpha}^{0}$ and $\theta_{\alpha}^{\,0}$ are constants. The term
denoted by $V_{\mathrm{ff}}^{\mathrm{tors}}(\phi_{\alpha})$ is a
commonly included torsional potential that depends only on the
dihedral angles $\phi_{\alpha}$ and
$V_{\mathrm{ff}}^{\mathrm{long-range}}(q^{a})$ normally comprises
long-range interactions such as Coulomb or van der Waals; hence, it
depends on the atomic positions $\vec{x}_{\alpha}^{\,\prime}$ which,
in turn, depend on all the internal coordinates $q^{a}$.

One of the reasons given for neglecting $\det \mathcal{H}$, when
classical force fields are used with potential energy functions such
as the one in eq.~\ref{eq:Vff}, is that the harmonic constraining
terms dominate over the rest of interactions and, since the constants
appearing on these terms (the $K_{r_{\alpha}}$, $K_{\theta_{\alpha}}$
in eq.~\ref{eq:Vff}) are independent of the conformation by
construction, so is $\det \mathcal{H}$
\cite{PE:Den2000MP,PE:Ber1983BOOK,PE:Go1976MM}. Here, we analyze a
more realistic quantum-mechanical potential and these considerations
are not applicable, however, \emph{they also should be checked in the
case of classical force fields}, since, for a potential energy such as
the one in eq.~\ref{eq:Vff}), the quantities $K_{r_{\alpha}}$ and
$K_{\theta_{\alpha}}$ are finite and the long-range terms will also
affect the Hessian at each point of the constrained hypersurface
$\Sigma$, rendering its determinant \emph{conformation-dependent}.

For the same reason, \emph{even in classical force fields, the
equilibrium values of the hard coordinates are not the constant
quantities $r_{\alpha}^{0}$ and $\theta_{\alpha}^{\,0}$} in
eq.~\ref{eq:Vff} but some functions $f^{I}(q^{i})$ of the soft
coordinates (see eq.~\ref{eq:real_constraints}). This fact,
recognized by some authors
\cite{PE:Che2005JCC,PE:Hes2002JCP,PE:Zho2000JCP,PE:Go1976MM}, provokes
that, if one chooses to assume approximation (iii) and the constants
$r_{\alpha}^{0}$ and $\theta_{\alpha}^{\,0}$ appearing in
eq.~\ref{eq:Vff} are designated as the equilibrium values, the
potential energy in $\Sigma$ may be heavily distorted, the cause being
simply that the long-range interactions between atoms separated by
three covalent bonds are not fully relaxed \cite{PE:Che2005JCC}. This
effect is probably larger if bond angles, and not only bond lengths,
are also constrained, which may partially explain the different
dynamical behaviour found in ref.~\citen{PE:VGu1982MM} when comparing
these types of constraints in Molecular Dynamics simulations. In
quantum mechanical calculations of small dipeptides, on the other
hand, the fact that the bond lengths and bond angles depend on the
Ramachandran angles $(\phi,\psi)$ has been pointed out by Sch\"affer
et al. \cite{PE:Sch1995JMS}. Therefore, approximation (iii), which is
very common in the literature
\cite{PE:All2005BOOK,PE:Mor2004ACP,PE:Pat2004JCP,PE:Sch2003MP,PE:Pas2002JCP,PE:Fre2002BOOK,PE:Den2000MP,PE:Den1998JCP,PE:Maz1998JPA,PE:Maz1996PRE,PE:Per1994JCC,PE:VGu1989BOOK,PE:Cic1986CPR,PE:Ber1984BOOK,PE:Ber1983BOOK,PE:VGu1982MM,PE:Ral1979JFM,PE:Fix1978JCP,PE:Go1976MM,PE:Fix1974PNAS},
should be critically analyzed in each particular case.

Apart from the typical internal coordinates $q^{a}$ used until now, in
terms of which the constrained hypersurface $\Sigma$ is described by
the relations $q^{I}=f^{I}(q^{i})$ in eq.~\ref{eq:real_constraints},
with $I=M+7,\ldots,N$, one may define a different set $Q^{a}$ such
that, on $\Sigma$, the corresponding hard coordinates are arbitrary
constants $Q^{I}=C^{I}$ (the external coordinates $q^{A}$ and $Q^{A}$
are irrelevant for this part of the discussion). To do this, for
example, let

\begin{equation}
\label{eq:qtoQ}
\begin{array}{lll}
Q^{i}:=q^{i} & i=7,\ldots,M+6 & \mathrm{and} \\
Q^{I}:=q^{I}-f^{I}(q^{i})+C^{I} & I=M+7,\ldots,N \ . & 
\end{array}
\end{equation}

Well then, while the relation between bond lengths, bond angles and
dihedral angles (the typical $q^{a}$ \cite{PE:Ech2006JCCa}) and the
Euclidean coordinates is straightforward and simple, the expression of
the transformation functions $Q^{a}(x^{\mu})$ needs the knowledge of
the $f^{I}$, which must be calculated numerically in most real
cases. This drastically reduce the practical use of the $Q^{a}$,
however, it is also true that they are conceptually appealing, since
they have a property that closely match our intuition about what the
soft and hard coordinates should be (namely, that the hard coordinates
$Q^{I}$ are constant on the relevant hypersurface $\Sigma$); and this
is why we term them \emph{exactly separable hard and soft
coordinates}. Now, we must also point out that, although the real
internal coordinates $q^{a}$ do not have this property, they are
usually close to it. The customary labeling of soft and hard
coordinates in the literature is based on this circumstance. Somehow,
the dihedral angles are the ``softest'' of the internal coordinates,
i.e., the ones that ``vary the most'' when the system visits different
regions of the hypersurface $\Sigma$; and this is why we term the real
$q^{a}$ \emph{approximately separable hard and soft coordinates},
considering approximation (iii) as a useful reference case.

To sum up, the three simplifying assumptions (i), (ii) and (iii) in
the beginning of this section should be regarded as approximations in
the case of classical force fields, as well as in the case of the more
realistic quantum-mechanical potential investigated in this work, and
they should be critically assessed in the systems of interest. Here,
while studying the model dipeptide HCO-{\small L}-Ala-NH$_2$, no
simplifying assumptions of this type have been made.

\bibliography{constrained}

\begin{thebibliography}{10}

\bibitem{PE:Alo2004BOOK}
J.~L. Alonso, G.~A. Chass, I.~G. Csizmadia, P.~Echenique, and A.~Taranc{\'o}n.
\newblock Do theoretical physicists care about the protein folding problem?
\newblock In R.~F. {\'A}lvarez-Estrada et~al., editors, {\em Meeting on
  Fundamental Physics `Alberto Galindo'}. Aula Documental, Madrid, 2004.
\newblock (arXiv:q-bio.BM/0407024).

\bibitem{PE:Dob2003NAT}
C.~M. Dobson.
\newblock Protein folding and misfolding.
\newblock {\em Nature}, 426:884--890, 2003.

\bibitem{PE:Dil1999PSC}
K.~A. Dill.
\newblock Polymer principles and protein folding.
\newblock {\em Prot. Sci.}, 8:1166--1180, 1999.

\bibitem{PE:He1998JCPb}
S.~He and H.~A. Scheraga.
\newblock Brownian dynamics simulations of protein folding.
\newblock {\em J. Chem. Phys.}, 108:287, 1998.

\bibitem{PE:Aba1994JCC}
R.~A. Abagyan, M.~M. Totrov, and D.~A. Kuznetsov.
\newblock {ICM}: {A} new method for protein modeling and design: Applications
  to docking and structure prediction from the distorted native conformation.
\newblock {\em J. Comp. Chem.}, 15:488--506, 1994.

\bibitem{PE:VGu1982MM}
W.~F. Van~Gunsteren and M.~Karplus.
\newblock Effects of constraints on the dynamics of macromolecules.
\newblock {\em Macromolecules}, 15:1528--1544, 1982.

\bibitem{PE:Lev1969PROC}
C.~Levinthal.
\newblock How to fold graciously.
\newblock In J.~T.~P. DeBrunner and E.~Munck, editors, {\em Mossbauer
  Spectroscopy in Biological Systems}, pages 22--24, Allerton House,
  Monticello, Illinois, 1969. University of Illinois Press.

\bibitem{PE:Chu2000JCC}
H.~M. Chun, C.~E. Padilla, D.~N. Chin, M.~Watanabe, V.~I. Karlov, H.~E. Alper,
  K.~Soosaar, K.~B. Blair, O.~M. Becker, L.~S.~D. Caves, R.~Nagle, D.~N. Haney,
  and B.~L. Farmer.
\newblock {MBO(N)D}: {A} multibody method for long-time {M}olecular {D}ynamics
  simulations.
\newblock {\em J. Comp. Chem.}, 21:159--184, 2000.

\bibitem{PE:Rei2000PD}
S.~Reich.
\newblock Smoothed {L}angevin dynamics of highly oscillatory systems.
\newblock {\em Physica D}, 118:210--224, 2000.

\bibitem{PE:Rei1999JCOP}
S.~Reich.
\newblock Multiple time scales in classical and quantum-classical molecular
  dynamics.
\newblock {\em J. Comput. Phys.}, 151:49--73, 1999.

\bibitem{PE:Sch1997ARBBS}
T.~Schlick, E.~Barth, and M.~Mandziuk.
\newblock Biomolecular dynamics at long timesteps: {B}ridging the timescale gap
  between simulation and experimentation.
\newblock {\em Annu. Rev. Biophys. Biomol. Struct.}, 26:181--222, 1997.

\bibitem{PE:VKa1984AJP}
N.~G. Van~Kampen and J.~J. Lodder.
\newblock Constraints.
\newblock {\em Am. J. Phys.}, 52:419--424, 1984.

\bibitem{PE:Ral1979JFM}
J.~M. Rallison.
\newblock The role of rigidity constraints in the rheology of dilute polymer
  solutions.
\newblock {\em J. Fluid Mech.}, 93:251--279, 1979.

\bibitem{PE:Hel1979JCP}
E.~Helfand.
\newblock Flexible vs. rigid constraints in {S}tatistical {M}echanics.
\newblock {\em J. Chem. Phys.}, 71:5000, 1979.

\bibitem{PE:Go1976MM}
N.~G{\={o}} and H.~A. Scheraga.
\newblock On the use of classical statistical mechanics in the treatment of
  polymer chain conformation.
\newblock {\em Macromolecules}, 9:535, 1976.

\bibitem{PE:Fix1974PNAS}
M.~Fixman.
\newblock Classical {S}tatistical {M}echanics of constraints: {A} theorem and
  application to polymers.
\newblock {\em Proc. Natl. Acad. Sci. USA}, 71:3050--3053, 1974.

\bibitem{PE:Go1969JCP}
N.~G{\={o}} and H.~A. Scheraga.
\newblock Analysis of the contributions of internal vibrations to the
  statistical weights of equilibrium conformations of macromolecules.
\newblock {\em J. Chem. Phys.}, 51:4751, 1969.

\bibitem{PE:Mor2004ACP}
D.~C. Morse.
\newblock Theory of constrained {B}rownian motion.
\newblock {\em Adv. Chem. Phys.}, 128:65--189, 2004.

\bibitem{PE:Den2000MP}
W.~K. Den~Otter and W.~J. Briels.
\newblock Free energy from molecular dynamics with multiple constraints.
\newblock {\em Mol. Phys.}, 98:773--781, 2000.

\bibitem{PE:Pat2004JCP}
A.~Patriciu, G.~S. Chirikjian, and R.~V. Pappu.
\newblock Analysis of the conformational dependence of mass-metric tensor
  determinants in serial polymers with constraints.
\newblock {\em J. Chem. Phys.}, 121:12708--12720, 2004.

\bibitem{PE:Per1985MM}
D.~Perchak, J.~Skolnick, and R.~Yaris.
\newblock Dynamics of rigid and flexible constraints for polymers. {E}ffect of
  the {F}ixman potential.
\newblock {\em Macromolecules}, 18:519--525, 1985.

\bibitem{PE:Pea1979JCP}
M.~R. Pear and J.~H. Weiner.
\newblock Brownian dynamics study of a polymer chain of linked rigid bodies.
\newblock {\em J. Chem. Phys.}, 71:212, 1979.

\bibitem{PE:Cha1979JCP}
D.~Chandler and B.~J. Berne.
\newblock Comment on the role of constraints on the conformational structure of
  n-butane in liquid solvent.
\newblock {\em J. Chem. Phys.}, 71:5386--5387, 1979.

\bibitem{PE:Got1976JCP}
M.~Gottlieb and R.~B. Bird.
\newblock A {M}olecular {D}ynamics calculation to confirm the incorrectness of
  the random-walk distribution for describing the {K}ramers freely jointed
  bead-rod chain.
\newblock {\em J. Chem. Phys.}, 65:2467, 1976.

\bibitem{PE:Zho2000JCP}
J.~Zhou, S.~Reich, and B.~R. Brooks.
\newblock Elastic molecular dynamics with self-consistent flexible constraints.
\newblock {\em J. Chem. Phys.}, 111:7919, 2000.

\bibitem{PE:Ber1983BOOK}
H.~J.~C. Berendsen and W.~F. Van~Gunsteren.
\newblock Molecular {D}ynamics with constraints.
\newblock In J.~W. Perram, editor, {\em The Physics of Superionic Conductors
  and Electrode Materials}, volume {NATO ASI Series B92}, pages 221--240.
  Plenum Press, 1983.

\bibitem{PE:Mac1998BOOK}
A.~D. MacKerell~Jr., B.~Brooks, C.~L. Brooks~III, L.~Nilsson, B.~Roux, Y.~Won,
  and M.~Karplus.
\newblock {CHARMM}: The energy function and its parameterization with an
  overview of the program.
\newblock In P.~v.~R. Schleyer et~al., editors, {\em The Encyclopedia of
  Computational Chemistry}, pages 217--277. John Wiley \& Sons, Chichester,
  1998.

\bibitem{PE:Bro1983JCC}
B.~R. Brooks, R.~E. Bruccoleri, B.~D. Olafson, D.~J. States, S.~Swaminathan,
  and M.~Karplus.
\newblock {CHARMM}: A program for macromolecular energy, minimization, and
  dynamics calculations.
\newblock {\em J. Comp. Chem.}, 4:187--217, 1983.

\bibitem{PE:Cor1995JACS}
W.~D. Cornell, P.~Cieplak, C.~I. Bayly, I.~R. Gould, Jr. Merz, K.~M., D.~M.
  Ferguson, D.~C. Spellmeyer, T.~Fox, J.~W. Caldwell, and P.~A. Kollman.
\newblock A second generation force field for the simulation of proteins,
  nucleic acids, and organic molecules.
\newblock {\em J. Am. Chem. Soc.}, 117:5179--5197, 1995.

\bibitem{PE:Pea1995CPC}
D.~A. Pearlman, D.~A. Case, J.~W. Caldwell, W.~R. Ross, T.~E. Cheatham~III,
  S.~DeBolt, D.~Ferguson, G.~Seibel, and P.~Kollman.
\newblock {AMBER}, a computer program for applying molecular mechanics, normal
  mode analysis, molecular dynamics and free energy calculations to elucidate
  the structures and energies of molecules.
\newblock {\em Comp. Phys. Commun.}, 91:1--41, 1995.

\bibitem{PE:Jor1988JACS}
W.~L. Jorgensen and J.~Tirado-Rives.
\newblock The {OPLS} potential functions for proteins. {E}nergy minimization
  for crystals of cyclic peptides and {C}rambin.
\newblock {\em J. Am. Chem. Soc.}, 110:1657--1666, 1988.

\bibitem{PE:Jor1996JACS}
W.~L. Jorgensen, D.~S. Maxwell, and J.~Tirado-Rives.
\newblock Development and testing of the {OPLS} all-atom force field on
  conformational energetics and properties of organic liquids.
\newblock {\em J. Am. Chem. Soc.}, 118:11225--11236, 1996.

\bibitem{PE:Hal1996JCCa}
T.~A. Halgren.
\newblock Merck molecular force field. {I}. {B}asis, form, scope,
  parametrization, and performance of {MMFF94}.
\newblock {\em J. Comp. Chem.}, 17:490--519, 1996.

\bibitem{PE:Hal1996JCCb}
T.~A. Halgren.
\newblock Merck molecular force field. {II}. {MMFF94} van der {Waals} and
  electrostatica parameters for intermolecular interactions.
\newblock {\em J. Comp. Chem.}, 17:520--552, 1996.

\bibitem{PE:Hal1996JCCc}
T.~A. Halgren.
\newblock Merck molecular force field. {III}. {M}olecular geometrics and
  vibrational frequencies for {MMFF94}.
\newblock {\em J. Comp. Chem.}, 17:553--586, 1996.

\bibitem{PE:Hal1996JCCd}
T.~A. Halgren.
\newblock Merck molecular force field. {IV}. {C}onformational energies and
  geometries for {MMFF94}.
\newblock {\em J. Comp. Chem.}, 17:587--615, 1996.

\bibitem{PE:Hal1996JCCe}
T.~A. Halgren.
\newblock Merck molecular force field. {V}. {E}xtension of {MMFF94} using
  experimental data, additional computational data, and empirical rules.
\newblock {\em J. Comp. Chem.}, 17:616--641, 1996.

\bibitem{PE:Pas2002JCP}
M.~Pasquali and D.~C. Morse.
\newblock An efficient algorithm for metric correction forces in simulations of
  linear polymers with constrained bond lengths.
\newblock {\em J. Chem. Phys.}, 116:1834, 2002.

\bibitem{PE:Den1998JCP}
W.~K. Den~Otter and W.~J. Briels.
\newblock The calculation of free-energy differences by constrained
  molecular-dynamics simulations.
\newblock {\em J. Chem. Phys.}, 109:4139, 1998.

\bibitem{PE:Cic1986CPR}
G.~Ciccotti and J.~P. Ryckaert.
\newblock Molecular dynamics simulation of rigid molecules.
\newblock {\em Comput. Phys. Rep.}, 4:345--392, 1986.

\bibitem{PE:Ber1984BOOK}
H.~J.~C. Berendsen and W.~F. Van~Gunsteren.
\newblock Molecular {D}ynamics simulations: {T}echniques and approaches.
\newblock In A.~J. et~al. Barnes, editor, {\em Molecular Liquids-Dynamics and
  Interactions}, pages 475--500. Reidel Publishing Company, 1984.

\bibitem{PE:Fix1978JCP}
M.~Fixman.
\newblock Simulation of polymer dynamics. {I}. {G}eneral theory.
\newblock {\em J. Chem. Phys.}, 69:1527, 1978.

\bibitem{PE:Alv2002MP}
R.~F. {\'A}lvarez-Estrada and G.~F. Calvo.
\newblock Models for biopolymers based on quantum mechanics.
\newblock {\em Mol. Phys.}, 100:2957--2970, 2002.

\bibitem{PE:Alv2000MTS}
R.~F. {\'A}lvarez-Estrada.
\newblock Models of macromolecular chains based on {C}lassical and {Q}uantum
  {M}echanics: comparison with {G}aussian models.
\newblock {\em Macromol. Theory Simul.}, 9:83--114, 2000.

\bibitem{PE:Hes2002JCP}
B.~Hess, H.~Saint-Martin, and H.~J.~C. Berendsen.
\newblock Flexible constraints: {A}n adiabatic treatment of quantum degrees of
  freedom, with application to the flexible and polarizable mobile charge
  densities in harmonic oscillators model for water.
\newblock {\em J. Chem. Phys.}, 116:9602, 2002.

\bibitem{PE:Bor1927APL}
M.~Born and J.~R. Oppenheimer.
\newblock Zur {Q}uantentheorie der {M}olekeln.
\newblock {\em Ann. Phys. Leipzig}, 84:457--484, 1927.

\bibitem{PE:Wil1980BOOK}
E.~B. Wilson~Jr., J.~C. Decius, and P.~C. Cross.
\newblock {\em Molecular Vibrations: The Theory of Infrared and Raman
  Vibrational Spectra}.
\newblock Dover Publications, New York, 1980.

\bibitem{PE:Bar1995JCC}
E.~Barth, K.~Kuczera, B.~Leimkuhler, and R.~D. Skeel.
\newblock Algorithms for constrained {M}olecular {D}ynamics.
\newblock {\em J. Comp. Chem.}, 16:1192--1209, 1995.

\bibitem{PE:And1983JCOP}
H.~C. Andersen.
\newblock Rattle: {A} {``velocity''} version of the {S}hake algorithm for
  molecular dynamics calculations.
\newblock {\em J. Comput. Phys.}, 52:24--34, 1983.

\bibitem{PE:Ryc1977JCOP}
J.~P. Ryckaert, G.~Ciccotti, and H.~J.~C. Berendsen.
\newblock Numerical integration of the {C}artesian equations of motion of a
  system with constraints: {M}olecular dynamics of n-alkanes.
\newblock {\em J. Comput. Phys.}, 23:327--341, 1977.

\bibitem{PE:Sch2003MP}
J.~Schlitter and M.~Kl{\"{a}}n.
\newblock The free energy of a reaction coordinate at multiple constraints: a
  concise formulation.
\newblock {\em Mol. Phys.}, 101:3439--3443, 2003.

\bibitem{PE:VGu1989BOOK}
W.~F. Van~Gunsteren.
\newblock Methods for calculation of free energies and binding constants:
  {S}uccesses and problems.
\newblock In W.~F. Van~Gunsteren and P.~K. Weiner, editors, {\em Computer
  Simulations of Biomolecular Systems}, pages 27--59. Escom science publishers,
  Netherlands, 1989.

\bibitem{PE:Din2000JCC}
A.~R. Dinner.
\newblock Local deformations of polymers with nonplanar rigid main-chain
  coordinates.
\newblock {\em J. Comp. Chem.}, 21:1132--1144, 2000.

\bibitem{PE:Sch1998JCP}
J.~Schofield and M.~A. Ratner.
\newblock Monte {C}arlo methods for short polypeptides.
\newblock {\em J. Chem. Phys.}, 109:9177, 1998.

\bibitem{PE:Per1994JCC}
A.~J. Pertsin, J.~Hahn, and H.~P. Grossmann.
\newblock Incorporation of bond-lengths constraints in {M}onte {C}arlo
  simulations of cyclic and linear molecules: conformational sampling for
  cyclic alkanes as test systems.
\newblock {\em J. Comp. Chem.}, 15:1121--1126, 1994.

\bibitem{PE:Kna1993JCC}
E.~W. Knapp and A.~Irgens-Defregger.
\newblock Off-lattice {M}onte {C}arlo method with constraints: {L}ong-time
  dynamics of a protein model without nonbonded interactions.
\newblock {\em J. Fluid Mech.}, 14:19--29, 1993.

\bibitem{PE:Alm1990MP}
N.~G. Almarza, E.~Enciso, J.~Alonso, F.~J. Bermejo, and M.~{\'A}lvarez.
\newblock {M}onte {C}arlo simulations of liquid n-butane.
\newblock {\em Mol. Phys.}, 70:485--504, 1990.

\bibitem{PE:Ech2006JCCc}
P.~Echenique and I.~Calvo.
\newblock Explicit factorization of external coordinates in constrained
  {S}tatistical {M}echanics models.
\newblock {\it To appear in J. Comp. Chem.}, 2006.
\newblock (arXiv:q-bio.QM/0512033).

\bibitem{PE:Ech2006JCCa}
P.~Echenique and J.~L. Alonso.
\newblock Definition of {S}ystematic, {A}pproximately {S}eparable and {M}odular
  {I}nternal {C}oordinates {(SASMIC)} for macromolecular simulation.
\newblock {\it To be published in J. Comp. Chem.}, 2006.
\newblock (arXiv:q-bio.BM/0511004).

\bibitem{PE:Lan2005PSFB}
A.~L{\'{a}}ng, I.~G. Csizmadia, and A.~Perczel.
\newblock Peptide models. {XLV}: {C}onformational properties of
  {N-formyl-{\small L}-methioninamide} ant its relevance to methionine in
  proteins.
\newblock {\em PROTEINS: Struct. Funct. Bioinf.}, 58:571--588, 2005.

\bibitem{PE:Per2003JCC}
A.~Perczel, O.~Farkas, I.~Jakli, I.~A. Topol, and I.~G. Csizmadia.
\newblock Peptide models. {XXXIII}. {E}xtrapolation of low-level
  {H}artree-{F}ock data of peptide conformation to large basis set {SCF},
  {MP2}, {DFT} and {CCSD(T)} results. {T}he {R}amachandran surface of alanine
  dipeptide computed at various levels of theory.
\newblock {\em J. Comp. Chem.}, 24:1026--1042, 2003.

\bibitem{PE:Var2002JPCA}
R.~Vargas, J.~Garza, B.~P. Hay, and D.~A. Dixon.
\newblock Conformational study of the alanine dipeptide at the {MP2} and {DFT}
  levels.
\newblock {\em J. Phys. Chem. A}, 106:3213--3218, 2002.

\bibitem{PE:Yu2001JMS}
C.-H. Yu, M.~A. Norman, L.~Sch{\"{a}}fer, M.~Ramek, A.~Peeters, and C.~van
  Alsenoy.
\newblock Ab initio conformational analysis of {N}-formyl {L}-alanine amide
  including electron correlation.
\newblock {\em J. Mol. Struct.}, 567--568:361--374, 2001.

\bibitem{PE:Csa1999PBMB}
A.~G. Cs{\'{a}}sz{\'{a}}r and A.~Perczel.
\newblock Ab initio characterization of building units in peptides and
  proteins.
\newblock {\em Prog. Biophys. Mol. Biol.}, 71:243--309, 1999.

\bibitem{PE:All2005BOOK}
M.~P. Allen and D.~J. Tildesley.
\newblock {\em Computer simulation of liquids}.
\newblock Clarendon Press, Oxford, 2005.

\bibitem{PE:Fre2002BOOK}
D.~Frenkel and Smit B.
\newblock {\em Understanding molecular simulations: From algorithms to
  applications}.
\newblock Academic Press, Orlando FL, 2nd edition, 2002.

\bibitem{PE:Kar1981MM}
M.~Karplus and J.~N. Kushick.
\newblock Method for estimating the configurational entropy of macromolecules.
\newblock {\em Macromolecules}, 14:325--332, 1981.

\bibitem{PE:Mac2004JCC}
A.~R. MacKerell~Jr., M.~Feig, and C.~L. Brooks~III.
\newblock Extending the treatment of backbone energetics in protein force
  fields: {L}imitations of gas-phase quantum mechanics in reproducing protein
  conformational distributions in molecular dynamics simulations.
\newblock {\em J. Comp. Chem.}, 25:1400--1415, 2004.

\bibitem{PE:Bor2003JPCB}
A.~J. Bordner, C.~N. Cavasotto, and R.~A. Abagyan.
\newblock Direct derivation of van der {W}aals force fields parameters from
  quantum mechanical interaction energies.
\newblock {\em J. Phys. Chem. B}, 107:9601--9609, 2003.

\bibitem{PE:Bea1997JACS}
M.~Beachy, D.~Chasman, R.~Murphy, T.~Halgren, and R.~Friesner.
\newblock Accurate ab initio quantum chemical determination of the relative
  energetics of peptide conformations and assessment of empirical force fields.
\newblock {\em J. Am. Chem. Soc.}, 119:5908--5920, 1997.

\bibitem{PE:Arn1989BOOK}
V.~I. Arnold.
\newblock {\em Mathematical Methods of Classical Mechanics}.
\newblock Graduate Texts in Mathematics. Springer, New York, 2nd edition, 1989.

\bibitem{PE:Vis2006JPCA}
B.~Viskolcz, S.~N. Fejer, and I.~G. Csizmadia.
\newblock Thermodynamic functions of conformational changes. 2.
  {C}onformational entropy as a measure of information accumulation.
\newblock {\it To be published in J. Phys. Chem. A.}, 2006.
\newblock (ASAP article {\texttt
  http://pubs.acs.org/cgi-bin/abstract.cgi/jpcafh/asap/abs/jp058219t.html}).

\bibitem{PE:And2001JCP}
I.~Andricioaei and M.~Karplus.
\newblock On the calculation of entropy from covariance matrices of the atomic
  fluctuations.
\newblock {\em J. Chem. Phys.}, 115:6289--6292, 2001.

\bibitem{PE:Vol1959BOOKpp}
M.~V. Volkenstein.
\newblock {\em Configurational Statistical of Polymeric Chains}.
\newblock Interscience, New York, 1959.

\bibitem{PE:Ram1963JMB}
G.~N. Ramachandran and C.~Ramakrishnan.
\newblock Stereochemistry of polypeptide chain configurations.
\newblock {\em J. Mol. Biol.}, 7:95--99, 1963.

\bibitem{PE:Hni2003JCC}
V.~Hnizdo, A.~Fedorowicz, H.~Singh, and E.~Demchuk.
\newblock Statistical thermodynamics of internal rotation in a hindering
  potential of mean force obtained from computer simulations.
\newblock {\em J. Comp. Chem.}, 24:1172--1183, 2003.

\bibitem{PE:Dem2001MP}
E.~Demchuk and H.~Singh.
\newblock Statistical thermodynamics of hindered rotation from computer
  simulations.
\newblock {\em Mol. Phys.}, 99:627--636, 2001.

\bibitem{PE:Mar2000BOOKb}
K.~V. Mardia and P.~E. Jupp.
\newblock {\em Directional Statistics}.
\newblock John Wiley \& Sons, Chichester, 2000.

\bibitem{PE:Sch1993JCC}
M.~W. Schmidt, K.~K. Baldridge, J.~A. Boatz, S.~T. Elbert, M.~S. Gordon, H.~J.
  Jensen, S.~Koseki, N.~Matsunaga, K.~A. Nguyen, S.~Su, T.~L. Windus,
  M.~Dupuis, and J.~A. Montgomery.
\newblock {G}eneral {A}tomic and {M}olecular {E}lectronic {S}tructure {S}ystem.
\newblock {\em J. Comp. Chem.}, 14:1347--1363, 1993.

\bibitem{PE:Bak1996JCP}
J.~Baker, A.~Kessi, and B.~Delley.
\newblock The generation and use of delocalized internal coordinates in
  geometry optimization.
\newblock {\em J. Chem. Phys.}, 105:192--212, 1996.

\bibitem{PE:Day2003PRL}
I.~P. Daykov, T.~A. Arias, and T.~D. Engeness.
\newblock Robust ab initio calculation of condensed matter: Transparent
  convergence through semicardinal multiresolution analysis.
\newblock {\em Phys. Rev. Lett.}, 90:216402, 2003.

\bibitem{PE:Alo2006JCC}
J.~L. Alonso and P.~Echenique.
\newblock A physically meaningful method for the comparison of potential energy
  functions.
\newblock {\em J. Comp. Chem.}, 27:238--252, 2006.

\bibitem{PE:Dob1991BOOK}
J.~D. Dobson.
\newblock {\em Applied multivariate data analysis}, volume~I.
\newblock Springer-Verlag, New York, 1991.

\bibitem{PE:Lev1999BOOK}
I.~N. Levine.
\newblock {\em Quantum Chemistry}.
\newblock Prentice Hall, Upper Saddle River, 5th edition, 1999.

\bibitem{PE:Hal2001TCA}
M.~D. Halls, J.~Velkovski, and H.~B. Schlegel.
\newblock {Harmonic Frequency Scaling Factors for Hartree-Fock, S-VWN, B-LYP,
  B3-LYP, B3-PW91 and MP2 and the Sadlej pVTZ Electric Property Basis Set}.
\newblock {\em Theo. Chem. Acc.}, 105:413, 2001.

\bibitem{PE:Sco1996JPC}
A.~P. Scott and L.~Radom.
\newblock Harmonic vibrational frequencies: {A}n evaluation of
  {H}artree-{F}ock, {M{\o}ller-Plesset}, {Quadratic Configuration Interaction,
  Density Functional Theory}, and semiempirical scale factors.
\newblock {\em J. Phys. Chem.}, 100:16502--16513, 1996.

\bibitem{PE:Els2001CP}
M.~Elstner, K.~J. Jalkanen, M.~Knapp-Mohammady, and S.~Suhai.
\newblock Energetics and structure of glycine and alanine based model peptides:
  {Approximate SCC-DFTB, AM1 and PM3 methods in comparison with DFT, HF and
  MP2} calculations.
\newblock {\em Chem. Phys.}, 263:203--219, 2001.

\bibitem{PE:Les2001BOOK}
A.~M. Lesk.
\newblock {\em Introduction to Protein Architecture}.
\newblock Oxford University Press, Oxford, 2001.

\bibitem{PE:Cha2001PBMB}
P.~Chakrabarti and D.~Pal.
\newblock The interrelationships of side-chain and main-chain conformations in
  proteins.
\newblock {\em Prog. Biophys. Mol. Biol.}, 76:1--102, 2001.

\bibitem{PE:Ber2000NAR}
H.~M. Berman, J.~Westbrook, Z.~Feng, G.~Gilliland, T.~N. Bhat, H.~Weissig,
  I.~N. Shindyalov, and P.~E. Bourne.
\newblock The protein data bank.
\newblock {\em Nucleic Acids Research}, 28:235--242, 2000.

\bibitem{PE:Gun1996JMB}
K.~Gunasekaran, C.~Ramakrishnan, and P.~Balaram.
\newblock Disallowed {R}amachandran conformations of amino acid residues in
  protein structures.
\newblock {\em J. Mol. Biol.}, 264:191--198, 1996.

\bibitem{PE:Cre1992BOOK}
T.~E. Creighton.
\newblock {\em Proteins: Structures and Molecular Properties}.
\newblock Freeman, W. H., New York, 2nd edition, 1992.

\bibitem{PE:Che2005JCC}
J.~Chen, W.~Im, and C.~L. Brooks~III.
\newblock Application of torsion angle molecular dynamics for efficient
  sampling of protein conformations.
\newblock {\em J. Comp. Chem.}, 26:1565--1578, 2005.

\bibitem{PE:Sch1995JMS}
L.~Sch{\"{a}}fer and C.~Ming.
\newblock Predictions of protein backbone bond distances and angles from first
  principles.
\newblock {\em J. Mol. Struct.}, 333:201--208, 1995.

\bibitem{PE:Maz1998JPA}
M.~Mazars.
\newblock Canonical partition function of freely jointed chains.
\newblock {\em J. Phys. A: Math. Gen.}, 31:1949--1964, 1998.

\bibitem{PE:Maz1996PRE}
M.~Mazars.
\newblock Statistical physics of the freely jointed chain.
\newblock {\em Phys. Rev. E}, 53:6297, 1996.

\end{thebibliography}

\end{document}